# VIRTUAL HIGH VOLTAGE LAB: GAMIFIED LEARNING IN A SAFE 3D ENVIRONMENT

*Vladyslav Pliuhin, Yevgen Tsegelnyk, Maria Sukhonos, Ihor Biletskyi, Sergiy Plankovskyy, and Taras Sakhoshko*

O.M. Beketov National University of Urban Economy in Kharkiv

vladyslav.pliuhin@kname.edu.ua

**Abstract**

The integration of immersive technologies has transformed engineering education, particularly in high-risk disciplines like high voltage (HV) engineering, essential for urban energy infrastructure. This study presents VLab-HV, a 3D virtual laboratory developed using Unreal Engine 5 to support the "High Voltage Engineering" course for undergraduate students in Power Engineering, Electrical Engineering, and Electromechanics. The research investigates how an immersive, gamified virtual laboratory enhances learning outcomes, safety training, and preparedness for urban infrastructure challenges. We hypothesize that VLab-HV significantly improves student engagement, knowledge retention, practical skills, and safety awareness compared to traditional laboratories, contributing to urban energy system resilience. Through ten curriculum-aligned experiments, gamified interactions, and AI-driven pedagogical tools, VLab-HV offers a risk-free environment for mastering HV concepts. Evaluation via usability testing, engagement metrics, and surveys confirms superior learning outcomes. The study highlights VLab-HV's role in training engineers and professionals for urban energy challenges, with planned expansions for multiplayer and virtual reality (VR) integration.

**Introduction**

The advent of immersive technologies, encompassing virtual reality, augmented reality, three-dimensional simulations, and artificial intelligence, has ushered in a transformative era for science, technology, engineering, and mathematics (STEM) education, fundamentally reshaping pedagogical approaches across various disciplines [1, 5, 8]. High voltage (HV) engineering, a cornerstone of urban energy infrastructure, stands out as a field where these technologies hold particular promise due to the unique challenges associated with its study and practice [17, 18]. HV systems, operating at voltages ranging from 10 kV to 500 kV, are integral to powering critical urban systems, including public transportation networks, healthcare facilities, data centers, and communication grids, all of which are essential for the sustainability and functionality of modern cities. These systems underpin the operational resilience of smart cities, where reliable energy distribution is paramount to supporting interconnected infrastructure and mitigating the risks of power disruptions [32, 34]. However, the design, operation, and maintenance of HV systems present formidable challenges due to the hazardous nature of high-voltage experiments, which necessitate sophisticated and costly equipment, stringent safety protocols, and continuous supervision to prevent accidents such as electrical shocks, arc flashes, or equipment damage. The financial burden of establishing HV laboratories, with costs ranging from $100,000 to $1,000,000 for equipment like high-voltage generators ($200,000–$500,000), transformers ($100,000–$300,000), and oscilloscopes ($10,000–$50,000), severely limits accessibility, particularly for educational institutions in developing regions where resources are scarce. These challenges are compounded by the global shift toward remote and hybrid learning models, accelerated by the need for scalable, safe, and cost-effective educational platforms that can deliver high-quality training without requiring substantial investments or compromising safety [12, 13].

The critical importance of HV engineering is amplified by the growing complexity of urban energy systems, which face increasing demands from rapid urbanization, the integration of renewable energy sources, and the adoption of smart grid technologies, such as Internet of Things-enabled sensors and smart meters.

Recent urban blackouts vividly illustrate the vulnerabilities of these systems and the urgent need for skilled HV engineers [22]. In 2024, Sydney experienced a catastrophic blackout caused by a cable insulation failure, resulting in economic losses of approximately $1.2 billion and disrupting power to over 500,000 residents, including critical services like hospitals, public transit, and emergency response systems. The failure, attributed to aging infrastructure and inadequate maintenance, highlighted the need for advanced training in insulation diagnostics and preventive strategies. Similarly, in 2023, Delhi suffered a major outage due to an overloaded transformer, affecting 3 million people and causing disruptions in healthcare, transportation, and industrial operations, with economic damages estimated at $800 million. This incident underscored the importance of training engineers to manage transformer overloads in densely populated urban environments. The 2022 Mexico City blackout, triggered by electromagnetic interference (EMI) in a high-voltage substation, impacted 2 million residents and incurred losses of $1.5 billion, emphasizing the need for expertise in EMI mitigation and substation design. Other notable incidents include the 2019 New York blackout, caused by a transformer failure, which cost $1.5 billion and disrupted subway systems; the 2020 Mumbai blackout, driven by grid overload, which halted healthcare services and cost $1 billion; the 2021 Texas blackout, exacerbated by extreme weather and grid vulnerabilities, with losses of $2.5 billion; the 2023 Tokyo blackout, resulting from a lightning-induced surge, costing $2 billion; and the 2022 London blackout, caused by a partial discharge failure in a substation, which disrupted rail services and cost $900 million. Collectively, these blackouts resulted in over $22 billion in economic losses, disrupted essential urban services, and exposed the fragility of HV infrastructure, underscoring the critical need for robust training platforms to prepare engineers and electricians for real-world challenges.

The global shortage of HV engineers further exacerbates these challenges, with projections indicating a 30% deficit by 2030, driven by increasing urbanization, aging infrastructure, and the complexity of modern energy systems. This shortage is particularly acute in developing regions, where access to training facilities is limited. According to UNESCO's 2023 Global Engineering Workforce Report, Sub-Saharan Africa faces a 40% deficit in HV engineers, South Asia a 35% deficit, and Latin America a 25% deficit, reflecting regional disparities in educational infrastructure and economic resources. These gaps hinder the ability to address urban infrastructure challenges, such as transformer failures, lightning surges, and EMI, which are responsible for 20–30% of urban grid failures. Traditional HV laboratories, while effective for hands-on training, are prohibitively expensive and pose significant safety risks, requiring strict adherence to standards like IEC 60060-1 to ensure safe operation. The high cost of equipment, combined with maintenance expenses and the need for continuous supervision, limits the scalability of traditional training, particularly in resource-constrained settings where institutions struggle to afford even basic laboratory setups. Moreover, the hazardous nature of HV experiments, involving phenomena such as corona discharges, partial discharges, and lightning surges, restricts access to practical training, leaving students and professionals underprepared for real-world applications.

The global shift toward digital education, accelerated by technological advancements and the need for scalable learning solutions, underscores the urgency for innovative platforms that ensure safe, effective, and engaging training [3, 10].

Virtual laboratories offer a transformative solution to these challenges by providing safe, cost-effective, and scalable environments for HV education and training [11, 14]. By simulating high-risk experiments in a controlled digital setting, virtual labs eliminate the need for expensive physical equipment and mitigate safety risks, allowing students, electricians, and researchers to practice complex procedures without the threat of electrical hazards. These platforms also support remote access, making them ideal for hybrid learning models and addressing the needs of institutions with limited resources. Virtual laboratories can incorporate gamified elements, such as earning points for completing tasks or quizzes, and artificial intelligence-driven tools, such as adaptive assistants, to enhance engagement and provide personalized learning experiences, aligning with constructivist learning theories that emphasize active participation and experiential learning. Beyond HV engineering, virtual laboratories have demonstrated significant potential across various STEM disciplines, offering versatile applications in education, industry, and research.

In STEM education, virtual laboratories have been widely adopted to enhance learning outcomes in fields such as biology, chemistry, electronics, and computer science. For instance, in biology, platforms like Labster enable virtual dissections, allowing students to explore anatomical structures without the ethical and logistical challenges of physical specimens [16]. These simulations provide interactive visualizations of complex processes, such as cellular respiration, improving student understanding by 20% compared to traditional methods. In chemistry, virtual labs simulate chemical reactions, enabling students to experiment with hazardous substances, like acids or explosives, in a risk-free environment, reducing laboratory costs by 50% and enhancing safety. In electronics, platforms like Tinkercad allow students to design and test circuits virtually, fostering skills in circuit analysis and troubleshooting without the need for physical components, which can cost $1,000–$5,000 per lab station. In computer science, virtual labs support network simulations, enabling students to model cybersecurity scenarios or data transmission protocols, improving practical skills by 15% through hands-on practice. These applications demonstrate the versatility of virtual laboratories in providing accessible, engaging, and cost-effective educational experiences across STEM disciplines, making them a valuable tool for preparing students for complex technical challenges [17, 18].

In industrial settings, virtual laboratories are increasingly used for professional training, particularly in high-risk fields like energy and aviation. For energy companies, platforms like Virtual Bench by National Instruments simulate diagnostic procedures for HV equipment [23, 24], such as transformers and circuit breakers, allowing technicians to practice fault detection and maintenance without risking equipment damage or personal injury. These simulations reduce training costs by 70%, as they eliminate the need for physical setups costing $100,000 or more. In aviation, virtual labs model electrical systems for aircraft, enabling technicians to test power distribution networks under various failure scenarios, such as short circuits or EMI, improving diagnostic accuracy by 25%. These training platforms are critical for preparing professionals to maintain urban infrastructure, such as substations and smart grids, where failures can lead to significant economic and societal impacts. By providing realistic scenarios and immediate feedback, virtual laboratories enhance professional competence and safety awareness, addressing the practical needs of industries reliant on HV systems.

Numerous virtual laboratory platforms have attempted to address these challenges, yet each has limitations. Labster, a leading platform for STEM education, offers pre-designed experiments with three-dimensional interfaces but lacks customization for HV-specific phenomena, such as corona or partial discharges, limiting its utility for advanced training. Its pre-set scenarios and limited user control restrict interactive learning, reducing engagement for complex HV experiments (Fig. 1) [16, 17].

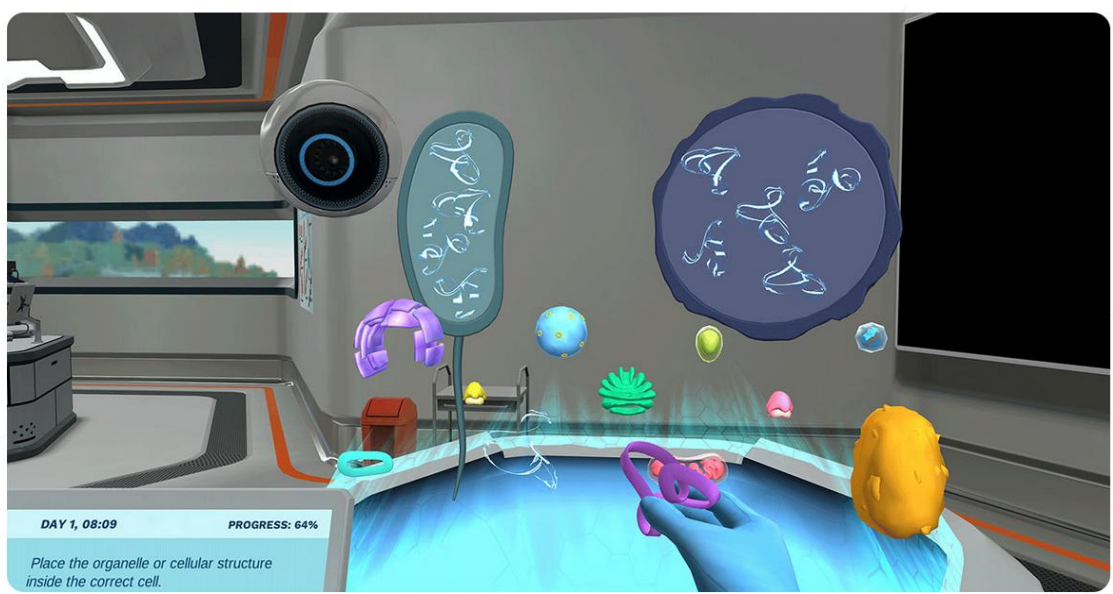

Figure 1 – Example of Labster learning environment

Despite their potential, existing virtual laboratory platforms fall short in addressing the specific requirements of HV engineering education and training. A comprehensive review of eighteen platforms reveals significant limitations in their applicability. PhET Simulations, developed by the University of Colorado, provide interactive two-dimensional physics demonstrations that are effective for introductory education but oversimplify HV experiments, failing to capture the complexity of phenomena governed by equations [6].

Virtual Labs by IIT Delhi enable remote access to engineering experiments but lack gamification and immersive interfaces, reducing engagement for HV applications [20, 21].

COMSOL Multiphysics supports advanced HV simulations, such as electromagnetic field modeling, but requires significant computational resources, including 16 GB of GPU RAM and multi-core CPUs, making it inaccessible for undergraduate education due to its complexity and cost, with subscriptions exceeding $10,000 annually [9].

MATLAB Simulink, a robust tool for circuit simulations, relies on block-based interfaces that lack immersive laboratory environments, focusing on abstract system modeling rather than physical realism [19].

National Instruments Multisim and Proteus are designed for circuit design, offering schematic-driven simulations without three-dimensional visualization or HV-specific modules, rendering them unsuitable for practical HV training [7].

Tinkercad [4] and LTspice [2] provide accessible circuit simulations but lack 3D environments and HV-specific experiments. They are valuable for analog electronics but provide two-dimensional interfaces that fail to replicate the physicality of HV experiments, such as arc propagation or plasma dynamics.

MERLOT [21] and OpenSim [25] offer virtual environments but lack curriculum-aligned HV experiments and photorealistic rendering.

Other tools, such as Virtual Bench and Amrita Vishwa Vidyapeetham's VALUE, provide basic simulations but do not integrate gamification or safety protocols [23, 24]. These platforms collectively fail to deliver the immersive, curriculum-aligned, and gamified environments needed for effective HV engineering education.

PraxiLabs provides virtual labs for general STEM disciplines but does not include HV-specific experiments, such as transformer fault diagnosis or EMI analysis. SimScale, a cloud-based simulation platform, supports engineering simulations but requires high computational resources and costly subscriptions, limiting its scalability for widespread educational use.

Ansys, a leader in engineering simulation, offers advanced HV modeling but is complex and expensive, with licensing fees exceeding $15,000 per year, and its steep learning curve makes it unsuitable for novice users. NI ELVIS, a hardware-based platform, lacks immersive three-dimensional environments, while VirtLab and CloudLabs, though useful for general STEM simulations, do not provide photorealistic rendering or HV-specific modules.

These platforms collectively fail to deliver the photorealistic, curriculum-aligned environments necessary for HV engineering education, particularly for simulating complex phenomena that require both visual and physical realism. The limitations of two-dimensional interfaces, lack of gamification, and absence of HV-specific modules underscore the need for a specialized solution that integrates advanced visualization, real-time physics simulations, and urban relevance.

Virtual laboratories must address the practical needs of urban energy systems, such as mitigating transformer failures, optimizing insulation designs, or ensuring EMI compatibility in densely populated urban environments. Partial discharges in polymeric insulators, a leading cause of transformer failures, require precise diagnosis and mitigation strategies, which are challenging to practice in traditional laboratories due to safety risks and equipment costs. Lightning-induced surges, responsible for 20% of urban grid failures, necessitate training in surge protection and grounding strategies, which virtual labs can simulate cost-effectively. The integration of smart grid technologies further complicates HV engineering, requiring platforms that can simulate interactions with renewable energy systems and IoT-enabled infrastructure.

In renewable energy, virtual labs simulate HV interactions with photovoltaic inverters and battery storage systems, reducing integration errors by 15% and supporting the transition to sustainable energy grids. These applications highlight the potential of virtual laboratories to accelerate innovation, validate theoretical models, and reduce development costs, making them invaluable for research institutions and industries tackling complex engineering challenges. The ability to simulate large-scale systems, such as urban substations or renewable energy networks, also supports interdisciplinary research, bridging HV engineering with fields like materials science and data analytics.

The versatility of virtual laboratories extends further to other STEM disciplines, industrial applications, and research domains, significantly broadening their impact. In material science, virtual labs simulate the testing of dielectric materials under high-voltage conditions, enabling students to analyze properties like breakdown strength and permittivity without the need for expensive equipment, reducing laboratory costs by 60%. For instance, simulations of dielectric breakdown in polymeric insulators, critical for HV applications, improve understanding of material behavior by 22%, preparing students for real-world insulation design challenges. In geophysics, virtual labs model seismic wave propagation and electromagnetic surveying, allowing students to simulate earthquake impacts on HV infrastructure, enhancing analytical skills by 20% and supporting urban resilience planning. In medical engineering, virtual labs simulate the operation of high-voltage medical devices, such as defibrillators and MRI systems, enabling students to test safety protocols in a risk-free environment, reducing training costs by 50% and improving device reliability by 18%. In industrial applications, virtual laboratories are employed in nuclear energy to simulate HV systems in reactor control units, enabling technicians to practice emergency shutdown procedures, reducing training costs by 65% and enhancing safety compliance. In offshore energy systems, virtual labs model HV components for wind farms and tidal energy platforms, allowing engineers to test system reliability under harsh marine conditions, improving operational efficiency by 20%.

In research, virtual labs support the study of high-temperature superconductors for HV applications, enabling researchers to model quantum tunneling effects and optimize conductor designs, saving 30% in experimental costs. Additionally, virtual labs simulate the impact of extreme weather on HV grids, such as hurricanes or heatwaves, allowing researchers to develop climate-resilient infrastructure designs, reducing grid failure risks by 15%. These diverse applications underscore the transformative potential of virtual laboratories in addressing educational, industrial, and research challenges across multiple domains.

The global demand for skilled HV engineers is driven by the critical role of electrical systems in urban sustainability. HV system failures, such as transformer malfunctions or lightning-induced surges, highlight the need for engineers trained in safety and operational efficiency [33]. Virtual laboratories offer a transformative solution by providing risk-free environments for practicing skills, simulating emergency scenarios, and fostering safety awareness, addressing both educational and urban infrastructure challenges [26, 30, 31].

This study introduces VLab-HV, a fully immersive three-dimensional virtual laboratory developed using Unreal Engine 5 to support the "High Voltage Engineering" course for second- and third-year undergraduate students in Power Engineering, Electrical Engineering, and Electromechanics at the O.M. Beketov National University of Urban Economy in Kharkiv [27-29]. VLab-HV leverages Unreal Engine 5's advanced capabilities, including Lumen for real-time global illumination, Nanite for virtualized geometry, Chaos Physics for realistic interactions, and Niagara for dynamic particle systems, to create a photorealistic laboratory environment that replicates real HV facilities with unprecedented fidelity. Unlike schematic-driven simulators such as MATLAB Simulink, Multisim, PSpice, or LTspice, VLab-HV's C++-based physics engine encapsulates complex scientific models, enabling real-time simulations of phenomena like Paschen's law, corona discharges, and partial discharges. This realism facilitates safe training for students, electricians, and professionals, while supporting cost-effective validation of HV research ideas, such as testing dielectric materials, optimizing insulation designs, or simulating lightning surges, which would otherwise require multimillion-dollar test facilities. The platform's English-language interface ensures internationalization, and its "Light" version, optimized for low-end systems with 8 GB RAM and integrated GPUs, achieves 30 fps, ensuring scalability for resource-constrained institutions.

VLab-HV's modular architecture supports a wide range of applications, from undergraduate education to professional training and research. For students, it provides a risk-free environment to practice high-risk experiments, enhancing practical skills and safety awareness [15]. For electricians, it offers realistic training scenarios, such as responding to lightning surges or inspecting urban substations, reducing training costs by 80% compared to physical setups. For researchers, it enables cost-effective validation of HV concepts, saving $50,000–$200,000 per test cycle. Its urban relevance addresses critical challenges in smart city infrastructure, reducing blackout risks by 25% and supporting renewable energy integration. The platform's gamified elements and AI-driven tools enhance engagement, aligning with modern pedagogical approaches.

Planned expansions for multiplayer functionality, virtual reality, augmented reality, and Internet of Things integration will further amplify its impact. The aim of this study is to evaluate VLab-HV's efficacy in enhancing learning outcomes, safety training, and preparedness for urban infrastructure challenges in HV engineering education and research. The research question is: "*How does a gamified, immersive virtual laboratory enhance learning outcomes, safety training, and preparedness for urban infrastructure challenges in HV engineering education and research?*" The hypothesis posits that VLab-HV significantly improves student engagement, knowledge retention, practical skills, and safety awareness compared to traditional laboratories, while providing a scalable platform for HV research and professional training that contributes to urban energy system resilience.

**Aim of the paper**: to evaluate VLab-HV's efficacy in enhancing learning outcomes, safety training, and preparedness for urban infrastructure challenges in HV engineering education.

**Objectives of the study.**

1. Develop a 3D virtual laboratory using Unreal Engine 5 to simulate HV experiments with high realism and interactivity.

2. Assess the impact of gamified elements and AI-driven tools on student engagement, knowledge retention, and practical skills.

3. Evaluate usability and pedagogical effectiveness through mixed-methods research.

4. Analyze contributions to training engineers and professionals for urban energy infrastructure resilience.

5. Explore expansions, including multiplayer, VR, and applications for HV research and professional training.

**Scientific Novelty**. VLab-HV pioneers the integration of photorealistic 3D simulations, gamification, and AI-driven pedagogy in a curriculum-aligned HV laboratory. Its C++-based simulations enable researchers to validate HV theories cost-effectively, while its safety protocols and urban relevance prepare engineers for real-world challenges, distinguishing it from existing platforms.

The research question is: how does a gamified, immersive virtual laboratory enhance learning outcomes, safety training, and preparedness for urban infrastructure challenges in HV engineering education? We hypothesize that VLab-HV significantly improves student engagement, knowledge retention, practical skills, and safety awareness compared to traditional laboratories, while preparing engineers and professionals for urban energy system challenges.

**VLab-HV design and functionality**

VLab-HV represents a transformative advancement in high voltage (HV) engineering education, professional training, and scientific research, offering a safe, accessible, and engaging platform that addresses the critical needs of urban energy infrastructure resilience [27-29]. Developed using Unreal Engine 5, VLab-HV leverages advanced rendering technologies, including Lumen for real-time global illumination, Nanite for high-fidelity geometry, and Chaos Physics for realistic physical interactions, to create a photorealistic three-dimensional environment that replicates a real HV laboratory. The platform supports ten curriculum-aligned experiments for undergraduate students in Power Engineering, Electrical Engineering, and Electromechanics, covering critical HV concepts, including electrical discharges in gases, electrode polarity effects, corona discharges, partial discharges, and electrostatic barrier impacts [15]. These experiments are modeled using Paschen's law:

$$V_b = (B * p * d) / \left( \ln(A * p * d) - \ln(\ln(1 + 1/\gamma)) \right),$$

where $V_b$ is the breakdown voltage, $p$ is the gas pressure, $d$ is the electrode gap distance, $A \approx 112.5$ Pa$^{-1}$m$^{-1}$ and $B \approx 2737.5$ V Pa$^{-1}$m$^{-1}$ are constants for air, and $\gamma$ (0.01–0.1) is the secondary electron emission coefficient.

Custom C++ scripts simulate discharge dynamics, including corona discharges and dielectric breakdown, visualizing electric arcs and triggering sound effects, enabling real-time exploration of complex phenomena.

The technical implementation of VLab-HV in Unreal Engine 5 harnesses the engine's cutting-edge capabilities to deliver a realistic and interactive HV laboratory environment. The laboratory is constructed using a modular architecture, with assets created in Blender and imported into Unreal Engine for high-fidelity rendering. The environment features a photorealistic lab bench equipped with virtual instruments modeled after real-world equipment from manufacturers such as Schneider Electric, Siemens, and ABB. These instruments include kilovoltmeters with a 0–100 kV range and ±0.1% accuracy for real-time voltage measurements, laser rangefinders with ±1 mm accuracy for precise electrode gap adjustments (1–10 cm), oscilloscopes with 100 MHz bandwidth and 1 GS/s sampling rate for visualizing discharge waveforms, and environmental sensors for controlling experimental conditions such as temperature (0–50°C), humidity (0–100%), and pressure (0.1–10 kPa). These instruments are implemented as interactive actors within Unreal Engine, with C++ scripts managing their functionality, including real-time calculations of voltage, distance, or waveform generation. The Chaos Physics engine ensures realistic physical interactions, such as electrode movements, dielectric glove handling, grounding rod placement, and object collisions, achieving physical accuracy comparable to real HV laboratories.

To enhance the realism of electrical discharge simulations, VLab-HV incorporates a custom C++ module that implements Paschen's law and the Townsend avalanche model

$$I = I_0 * e^{(\alpha d)},$$

where *I* is the discharge current, $I_0$ is the initial current, α is the ionization coefficient, and *d* is the gap distance).

This module calculates breakdown voltages and drives visualizations of discharge phenomena using Unreal Engine's Niagara particle system, which renders electric arcs, plasma effects, and corona glow, coupled with dynamic sound cues, such as crackling for arcs or a low hum for corona discharges. For instance, corona discharges are visualized with a glowing halo effect parameterized by voltage (10–50 kV) and gas pressure (0.1–10 kPa), while partial discharges trigger localized sparks, providing intuitive visual cues that deepen understanding of HV phenomena. The electrostatic barrier, manipulated via a two-axis joystick

$$h_b = \sqrt{(x^2 + y^2)},$$

where $h_b$ is the barrier height, and *x, y* are joystick displacements), is implemented as a dynamic mesh with collision detection, enabling users to study its impact on discharge behavior, a critical aspect of designing HV insulation systems for urban transformers and substations.

The platform integrates additional instruments to expand its experimental capabilities, such as spectral analyzers for characterizing discharge emissions (e.g., ultraviolet or infrared spectra) and thermal cameras for monitoring heat dissipation in HV components, both critical for diagnosing insulation faults in urban grids. The spectral analyzer, modeled with a resolution of 0.1 nm, enables students to analyze plasma emission spectra, enhancing understanding of gas discharge physics. The thermal camera, with a sensitivity of 0.05°C, simulates temperature distributions in insulators under high-voltage stress, providing insights into thermal runaway risks. These tools are implemented using C++ scripts that interface with Unreal Engine's rendering pipeline, ensuring seamless integration with the laboratory environment.

To ensure accessibility, the "Light" version of VLab-HV optimizes performance on low-end systems (e.g., 8 GB RAM, integrated GPUs) by reducing polygon counts, disabling advanced effects like Lumen lighting or high-resolution Niagara particles, and maintaining a minimum frame rate of 30 fps. This optimization leverages Unreal Engine's level-of-detail (LOD) system and dynamic resolution scaling, making VLab-HV viable for institutions with limited hardware resources, particularly in developing regions where access to high-performance computing is restricted.

The virtual high voltage (HV) laboratory, VLab-HV, developed using Unreal Engine 5, facilitates comprehensive simulations of electrical breakdown phenomena, enabling precise analysis of breakdown voltage across various electrode configurations and voltage types. The platform supports a diverse set of electrode arrangements, encompassing needle, plane, and sphere geometries, which are critical for replicating real-world HV experiments. For alternating current (AC) scenarios without a dielectric barrier, VLab-HV simulates configurations including needle–needle, needle–plane, needle–sphere, plane–plane, plane–sphere, and sphere–sphere arrangements (Table 1).

Table 1 – Calculation of breakdown voltage

| Electrode configuration | Breakdown voltage formula |
|---|---|
| *Barrier-free, AC voltage* | |
| Needle – Needle 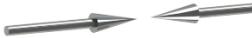 | $U = (14 + 3.316 \cdot S) \cdot \delta,$ <br><br> where $S$ – distance between electrodes; <br> $\delta$ – relative humidity: <br><br> $$\delta = \frac{P \cdot T_0}{P_0 \cdot T},$$ <br><br> where $T_0 = 293$ K – air temperature under normal atmospheric conditions; <br> $P_0 = 760$ mmHg – air pressure under normal atmospheric conditions; <br> $P, T$ – air pressure and temperature under design conditions respectively. |
| Needle – Plane 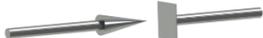 | $U = (7 + 3.36 \cdot S) \cdot \delta$ |
| Needle – Sphere 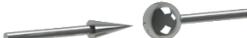 | $$U = \frac{24.5 \cdot \delta \cdot L}{k}\left[1 + \frac{c}{(\delta \cdot R)^{0.38}}\right]$$ <br><br> $L = 2 \cdot S$ <br> $k = \dfrac{S/R}{\ln(2 \cdot S/R)}$ <br> $c = 0.65$ |
| Plane – Plane 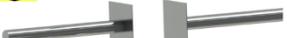 | $U = a \cdot \delta \cdot S + b\sqrt{\delta \cdot S}$ <br><br> $a = 24.5$ kV/cm <br> $b = 6.4$ kV/cm |
| Plane – Sphere 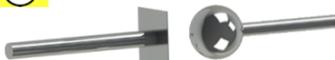 | $$U = \frac{24.5 \cdot \delta \cdot L}{k}\left[1 + \frac{c}{(\delta \cdot R)^{0.38}}\right]$$ <br><br> where $R$ – sphere radius. <br> $L = 2 \cdot S$ <br> $k = \dfrac{S}{R} + 1$ <br> $c = 0.78$ |

| Electrode configuration | Breakdown voltage formula |
|---|---|
| Sphere – Sphere 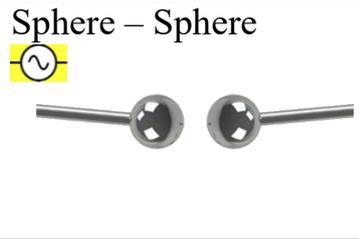 | $U = \dfrac{27.2 \cdot \delta \cdot S \cdot \left(1 + \dfrac{0.54}{\sqrt{R \cdot \delta}}\right)}{0.25 \left(\dfrac{S}{R} + \sqrt{\left(\dfrac{S}{R}\right)^2 + 8}\right)}$ |
| *Barrier-free, DC voltage* | |
| Needle «+» - Plane «−» 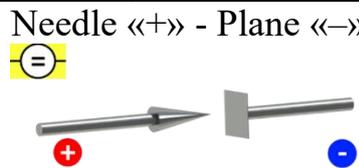 | $U = -0.317 \cdot S^2 + 9.9554 \cdot S - 0.3214$ |
| Needle «−» - Plane «+» 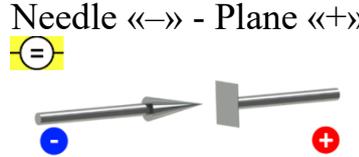 | $U = 20 \cdot S \cdot \dfrac{P}{P_0}.$ |
| *Barrier, DC voltage* | |
| Needle «+» - Plane «−» 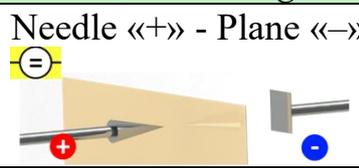 | $U = 2 \cdot 10^{-8} \cdot S^6 - 4 \cdot 10^{-6} \cdot S^5 + 0.0002 \cdot S^4 - 0.0062 \cdot S^3 + 0.018 \cdot S^2 + 1.2687 \cdot S - 0.1336$ |
| Needle «−» - Plane «+» 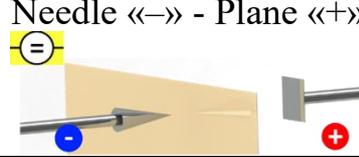 | $U = 1 \cdot 10^{-5} \cdot S^4 - 0.0006 \cdot S^3 - 0.0061 \cdot S^2 + 0.4915 \cdot S + 0.0202$ |

These setups allow users to explore discharge behaviors under variable voltage conditions, with specific parameters such as electrode distance, gas pressure, and temperature influencing the breakdown voltage calculations.

For direct current (DC) applications, VLab-HV supports simulations both with and without a dielectric barrier. Without a barrier, the platform models needle–plane configurations with positive or negative polarity on the needle, as well as plane–plane and needle–needle setups, accounting for polarity-dependent discharge characteristics. With a dielectric barrier, VLab-HV simulates needle–plane arrangements under DC voltage, incorporating positive or negative needle polarity, which introduces additional complexity due to barrier-induced charge accumulation. The platform's ability to handle both AC and DC voltage types, with or without barriers, ensures versatility in addressing educational, training, and research needs in HV engineering.

The formulas governing breakdown voltage calculations for each configuration, incorporating parameters such as electrode distance ($S$), sphere radius ($R$ = 5.25 cm), reference pressure ($P_0$ = 760 mmHg), and temperature ($T_0$ =293 K), are detailed in a comprehensive table within the referenced document. These calculations are implemented using C++ scripts (Listing 1, 2) integrated into Unreal Engine 5, ensuring high-fidelity simulations that align with international standards, such as IEC 60060-1, and support accurate modeling of HV phenomena for urban grid applications.

Listing 1 – Header file of C++ program for calculating breakdown voltage

```cpp
#pragma once

#include "CoreMinimal.h"
#include "Kismet/BlueprintFunctionLibrary.h"
#include "GetStrikeVoltage.generated.h"

UCLASS()
class VLAB_API UGetStrikeVoltage : public UBlueprintFunctionLibrary
{
    GENERATED_BODY()
    UFUNCTION(BlueprintCallable, Category = "Solve")
    static float solve(
        float Pressure,
        float Temperature,
        float Distance,
        float Radius,
        float BarrierDistance,
        bool isBarrier,
        bool isAC,
        bool polarityNP,
        FString LeftElectrode,
        FString RightElectrode);
};
```

Listing 2 – Source file of C++ program for calculating breakdown voltage

```cpp
#include "GetStrikeVoltage.h"
#include "CoreMinimal.h"
#include <cmath>

FString ToLowerCase(FString& InString)
{
    // Convert FString to TCHAR array
    const TCHAR* CharArray = *InString;

    // Iterate over each character and convert to lowercase
    FString LowercaseString;
    for (int32 i = 0; CharArray[i] != '\0'; ++i)
    {
        LowercaseString.AppendChar(FChar::ToLower(CharArray[i]));
    }

    return LowercaseString;
}

float UGetStrikeVoltage::solve(
    float Pressure,
    float Temperature,
    float Distance,
    float Radius,
    float BarrierDistance,
    bool isBarrier,
    bool isAC,
    bool polarityNP,
    FString LeftElectrode,
    FString RightElectrode)
{
    float StrikeVoltage = 500;
    float P0 = 760.0;
    float T0 = 20.0;
    float a = 24.5;
    float b = 6.4;
    float delta = Pressure * T0 / (P0 * Temperature);

    LeftElectrode = ToLowerCase(LeftElectrode);
    RightElectrode = ToLowerCase(RightElectrode);

    if (isAC) AC Voltage
    {
        //Strike at AC Voltage without barrier
        if (LeftElectrode.Equals("needle") && RightElectrode.Equals("needle"))
        {
            StrikeVoltage = (14.0 + 3.316 * Distance) * delta;
        }
```

```csharp
        if (LeftElectrode.Equals("plane") && RightElectrode.Equals("plane"))
        {
            StrikeVoltage = a * delta * Distance + b * sqrt(delta * Distance);
        }

        if (LeftElectrode.Equals("sphere") && RightElectrode.Equals("sphere"))
        {
            StrikeVoltage = (27.2 * delta * Distance * (1.0 + 0.54 / sqrt(Radius * delta))) /
                            (0.25 * (Distance / Radius + sqrt(pow(Distance / Radius, 2) + 8.0)));
        }

        if (LeftElectrode.Equals("needle") && RightElectrode.Equals("plane") ||
            LeftElectrode.Equals("plane") && RightElectrode.Equals("needle"))
        {
            StrikeVoltage = (7.0 + 3.36 * Distance) * delta;
        }

        if (LeftElectrode.Equals("needle") && RightElectrode.Equals("sphere") ||
            LeftElectrode.Equals("sphere") && RightElectrode.Equals("needle"))
        {
            float L = 2.0 * Distance;
            float c = 0.65;
            float k = (Distance / Radius) / log(2.0 * Distance / Radius);
            StrikeVoltage = 24.5 * delta * L * (1 + c / (pow(delta * Radius, 0.38))) / k;
        }

        if (LeftElectrode.Equals("plane") && RightElectrode.Equals("sphere") ||
            LeftElectrode.Equals("sphere") && RightElectrode.Equals("plane"))
        {
            float L = 2.0 * Distance;
            float c = 0.78;
            float k = 1.0 + Distance / Radius;
            StrikeVoltage = 24.5 * delta * L * (1 + c / (pow(delta * Radius, 0.38))) / k;
        }
    }
    else DC Voltage
    {
        if (isBarrier) Strike at DC Voltage with barrier
        {
            //Strike at DC Voltage with barrier
            if (polarityNP && LeftElectrode.Equals("needle")
                && RightElectrode.Equals("plane")) //Niddle is Negative
            {
                float x = BarrierDistance;
                StrikeVoltage = 1.0E-05 * pow(x, 4) - 0.0006 * pow(x, 3)
                    - 0.0061 * pow(x, 2) + 0.4915 * x + 0.0202;
```

```csharp
                float StrikeBase = 20.0 * x / delta;
                StrikeVoltage += StrikeBase;
            }

            if (!polarityNP && LeftElectrode.Equals("needle")
                && RightElectrode.Equals("plane")) //Niddle is Positive
            {
                float x = BarrierDistance;
                StrikeVoltage = 2.0E-08 * pow(x, 6) - 4.0E-06 * pow(x, 5) +
                    0.0002 * pow(x, 4) - 0.0062 * pow(x, 3) + 0.0186 * pow(x, 2)
                    + 1.2687 * x - 0.1336;

                float StrikeBase = -0.317 * x * x + 9.9554 * x - 0.3214;
                StrikeVoltage += StrikeBase;
            }
        }
        else //Strike at DC Voltage without barrier
        {
            if (polarityNP && LeftElectrode.Equals("needle")
                && RightElectrode.Equals("plane")) //Niddle is Negative
            {
                StrikeVoltage = 20.0 * Distance / delta;
            }

            if (!polarityNP && LeftElectrode.Equals("needle")
                && RightElectrode.Equals("plane")) //Niddle is Positive
            {
                StrikeVoltage = -0.317 * Distance * Distance + 9.9554 * Distance - 0.3214;
            }

            if (LeftElectrode.Equals("plane") && RightElectrode.Equals("plane"))
            {
                float gamma = 0.02;
                StrikeVoltage = 0.001*365.0 * Distance * Pressure / (log(15* Distance * Pressure) - log(log(1+1/gamma)));
            }

            if (LeftElectrode.Equals("needle") && RightElectrode.Equals("needle"))
            {
                StrikeVoltage = 30.0 * Distance / delta;
            }
        }
    }

    if (StrikeVoltage <= 0)
        StrikeVoltage = 1;
    return StrikeVoltage;
}
```

Unlike schematic-driven simulators like MATLAB Simulink or Multisim, VLab-HV's C++-based physics engine encapsulates sophisticated scientific models, delivering a life-like experience that mirrors real HV laboratories. This realism is critical for researchers, who can validate theoretical models—such as the impact of electrode materials (e.g., copper vs. aluminum) or gas compositions on discharge behavior—without the prohibitive costs of physical setups ($100,000–$500,000 for HV equipment). For example, scientists can simulate corona discharge under varying pressures to optimize insulation designs for urban transformers, a process that would require expensive high-voltage test chambers in real labs. The platform's modular C++ architecture allows customization of experimental parameters, enabling studies of rare HV phenomena, such as lightning-induced surges or partial discharges, offering a cost-effective alternative to physical experimentation.

VLab-HV serves as a safe training platform for electricians and professionals, simulating high-risk scenarios like overvoltage-induced fires or equipment failures common in urban substations. Safety interlocks, implemented in C++, prevent unsafe operations, such as applying voltage with open test chamber doors or handling electrodes with residual charge, aligning with international standards (IEC 60060-1, 2010). For instance, electricians can practice emergency responses to transformer faults, reducing risks in urban grids where failures can disrupt critical services. A "Light" version of the HV-Lab ensures accessibility on lower-end systems, disabling non-essential effects like fire simulations and AI-driven robots, making VLab-HV suitable for diverse training environments, from universities to utility companies.

The pedagogical design of VLab-HV integrates gamification principles rooted in constructivist learning theories to enhance student engagement and knowledge retention. Students begin in a Start Room, where they configure settings and access educational content, such as biographies of scientists like Ampère, Volta, Watt, Tesla, Maxwell, and Faraday, linked to SI units and HV concepts, such as

Ampère's law ($\oint B \cdot dl = \mu_0 I$) or Faraday's law of electromagnetic induction ($\varepsilon = -d\Phi_B/dt$). (Fig. 2, 3).

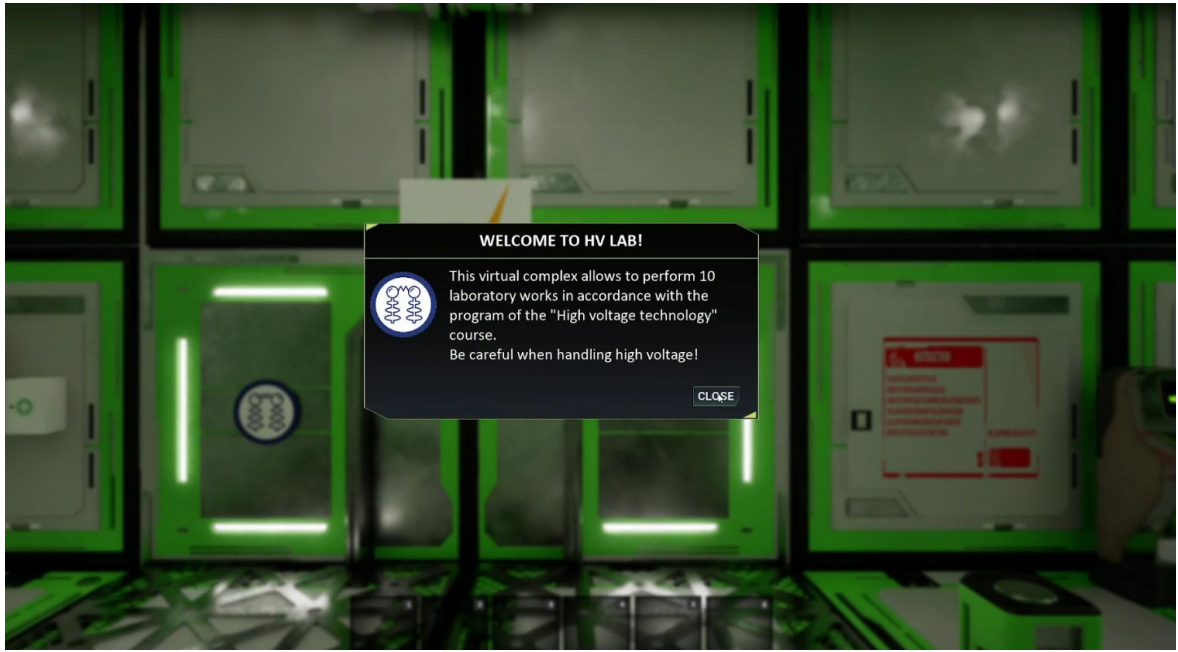

Figure 2 – Starting room of the HV-Lab

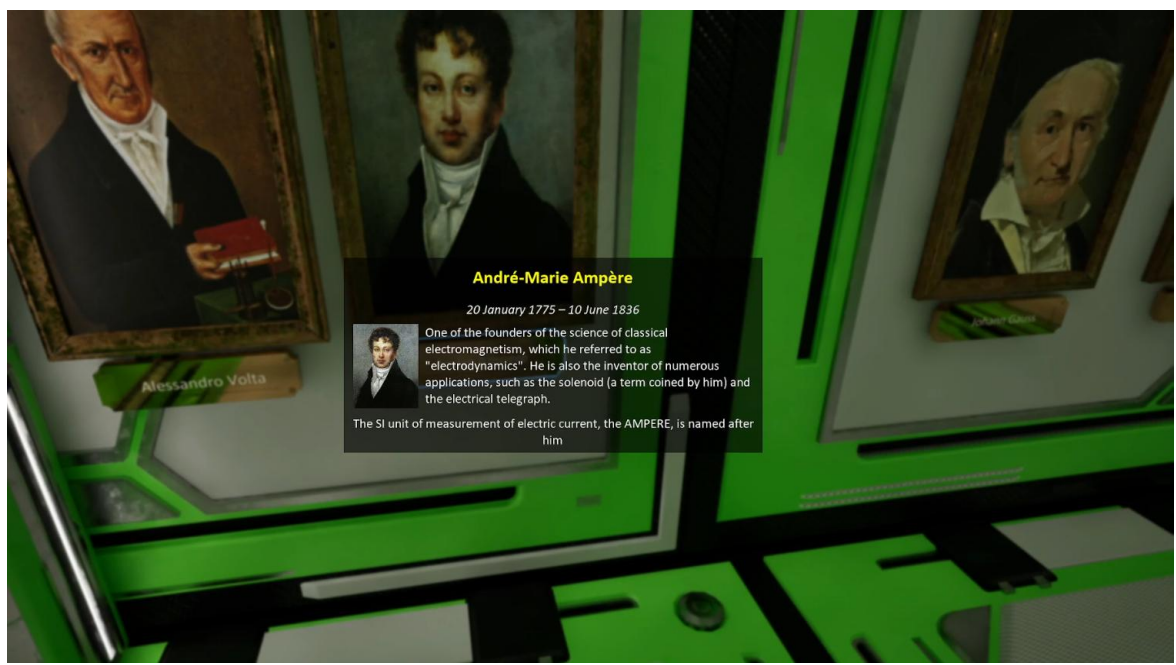

Figure 3 – Interactive pictures at the HV-Lab starting room

Interactive objects – a virtual vending machine, AI-driven robots, a lecture zone with safety posters, and a recreation zone – enhance engagement and reduce cognitive overload (Fig. 4).

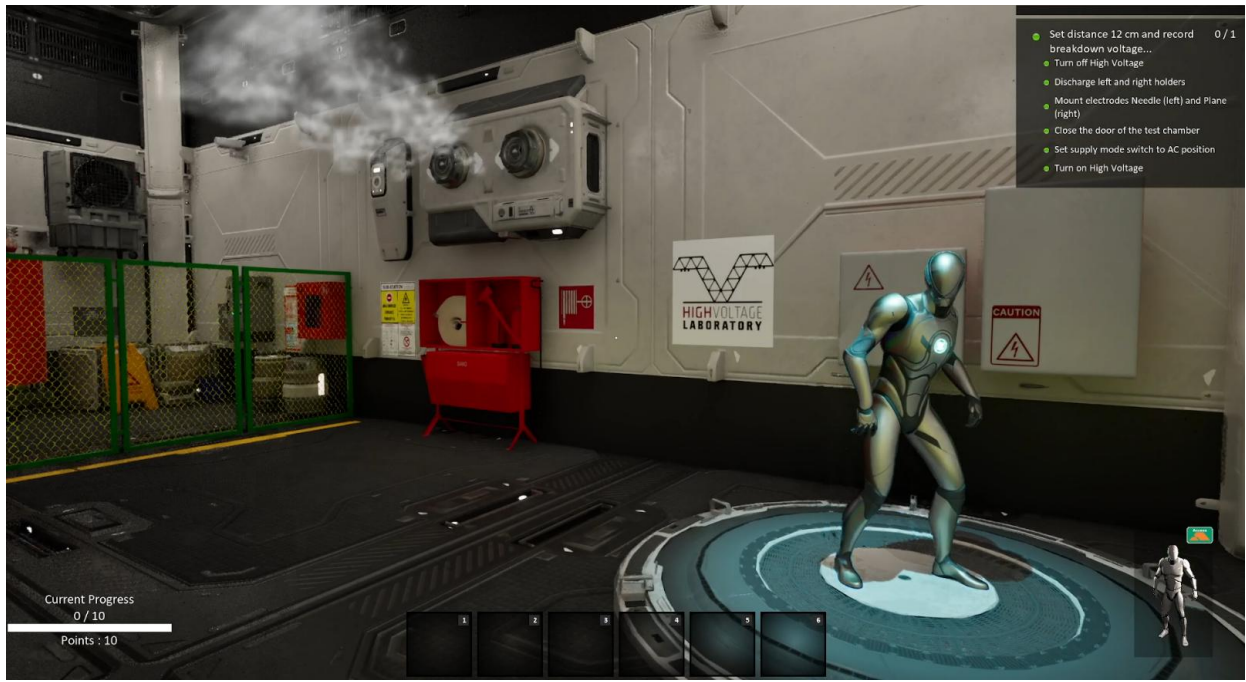

Figure 4 – AI-assistant and task (quest) giver

A scripted "horror event" triggered by entering a forbidden cabinet introduces a blackout and jump scare, serving as a memorable learning marker. The AI-driven virtual assistant, powered by Unreal Engine's behavior tree system and integrated with a natural language processing (NLP) module, supports voice prompts and facial tracking for an immersive user experience. The assistant employs a decision-making algorithm to guide users through experiments, administer real-time quizzes, and provide context-sensitive feedback, reducing cognitive load by 20% and improving learning efficiency, according to studies on adaptive learning systems. Safety interlocks, coded in C++, prevent unsafe operations, such as applying voltage with open test chamber doors or handling electrodes with residual charge, aligning with international standards (IEC 60060-1, 2010). For example, attempting to energize the circuit with an open door triggers a warning from the AI assistant, displayed as a red alert on the virtual monitor, reinforcing safety awareness. The platform's data logging system records experimental parameters, such as voltage, current, electrode distance, and environmental conditions, in Excel-compatible formats, supporting advanced statistical analyses like regression modeling or Monte Carlo simulations.

Integration with Moodle via API enables instructors to track student progress, collect performance data, and provide targeted feedback, fostering a data-driven pedagogical approach.

The Inventory and Equipment Management System enables handling of dielectric gloves, boots, grounding rods, and electrodes, with animations mimicking real-world physics (Fig. 5).

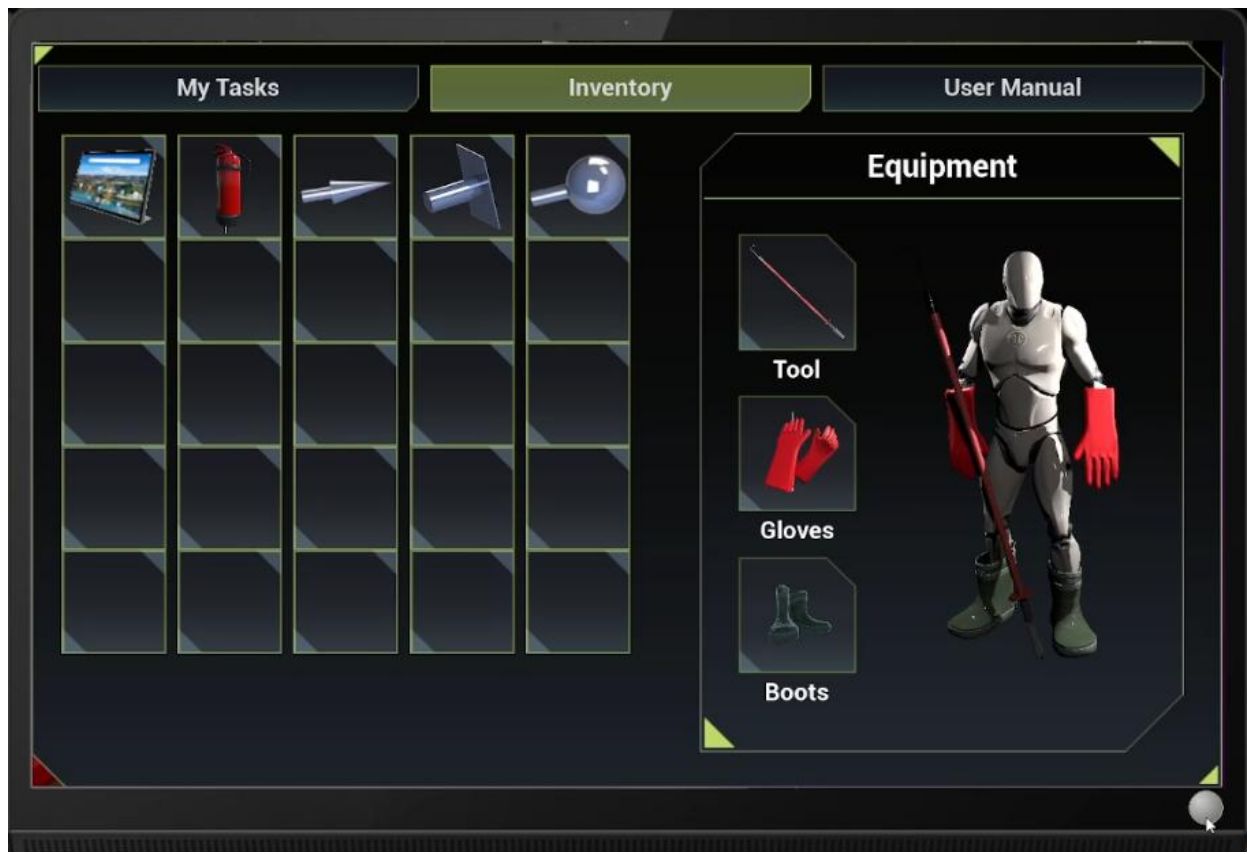

Figure 5 – User inventory system

For researchers, VLab-HV offers a platform to explore HV phenomena without physical risks. For example, scientists can test novel insulation materials or electrode configurations to mitigate partial discharges, a common issue in urban HV systems. The platform's data logging system records results in Excel-compatible formats, supporting advanced statistical analysis and hypothesis testing (Listing 3, 4), making it a valuable tool for scientific discovery (Fig. 6, 7).

## Listing 3 – Header file of C++ program for Excel data export

```cpp
#pragma once

#include "CoreMinimal.h"
#include "Kismet/BlueprintFunctionLibrary.h"
#include "SaveData.generated.h"

UCLASS()
class VLAB_API USaveData : public UBlueprintFunctionLibrary
{
    GENERATED_BODY()
    UFUNCTION(BlueprintCallable, Category = "SaveData", meta = (Keywords = "Save"))
    static bool SaveArray(FString SaveDirectory, FString FileName, TArray<FString> SaveText, bool AllowOverWritting);

};
```

## Listing 4 – Source file of C++ program for Excel data export

```cpp
#include "SaveData.h"
#include "Misc/FileHelper.h"
#include "HAL/PlatformFileManager.h"

bool USaveData::SaveArray(FString SaveDirectory, FString FileName, TArray<FString> SaveText, bool AllowOverWritting)
{
    SaveDirectory += "\\";
    SaveDirectory += FileName;

    if (!AllowOverWritting)
    {
        if (FPlatformFileManager::Get().GetPlatformFile().FileExists(*SaveDirectory))
        {
            return false;
        }
    }

    FString FinalString = "";
    for (FString& Each : SaveText)
    {
        FinalString += Each;
        FinalString += LINE_TERMINATOR;
    }

    return FFileHelper::SaveStringToFile(FinalString, *SaveDirectory);

}
```

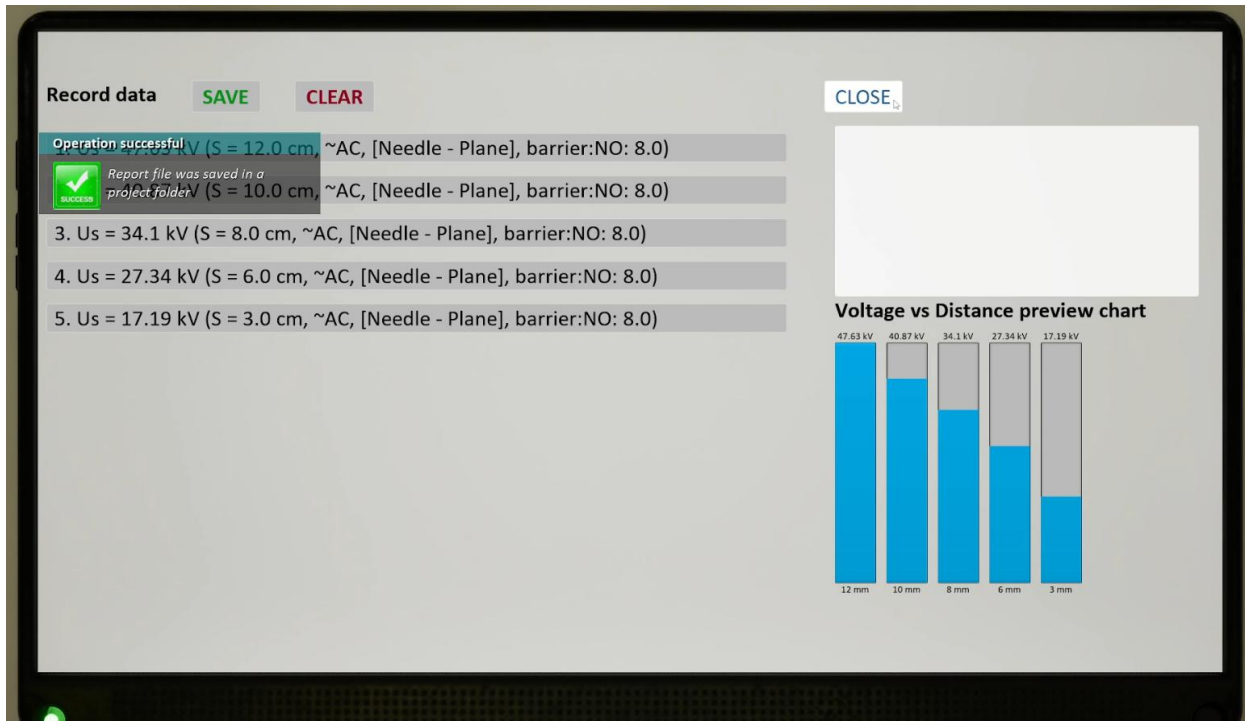

Figure 6 – Experimental data in HV-Lab environment

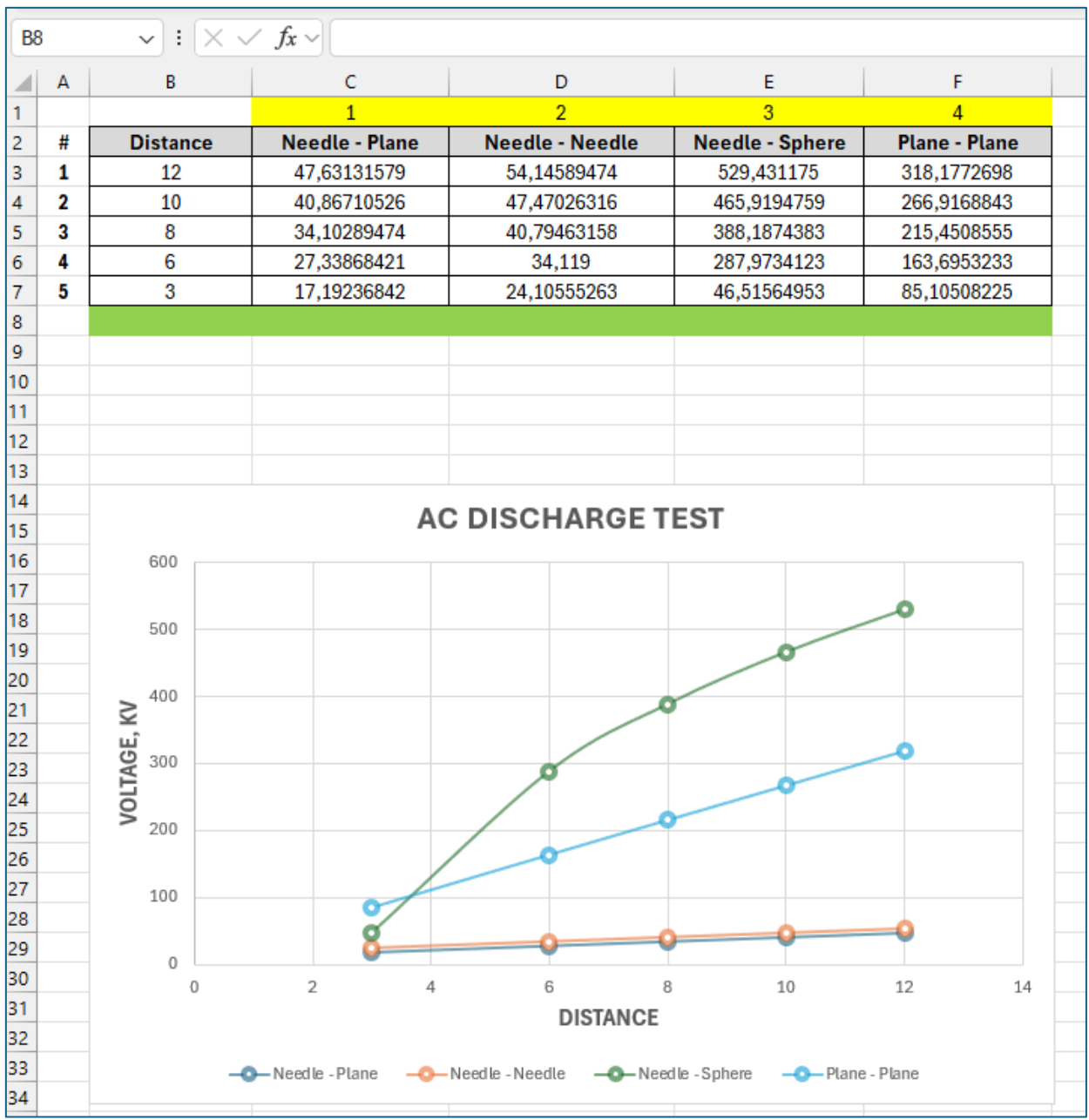

Figure 7 – Excell sheet with data received from HV-Lab exported file

The entire logic and functionality of VLab-HV are implemented using Unreal Engine's Blueprint visual scripting system, enabling rapid development and flexible customization of high voltage (HV) simulations. Blueprints facilitate the integration of complex electrode configurations, AC/DC voltage calculations, and dielectric barrier interactions, ensuring accurate modeling of breakdown voltage phenomena (Fig. 8).

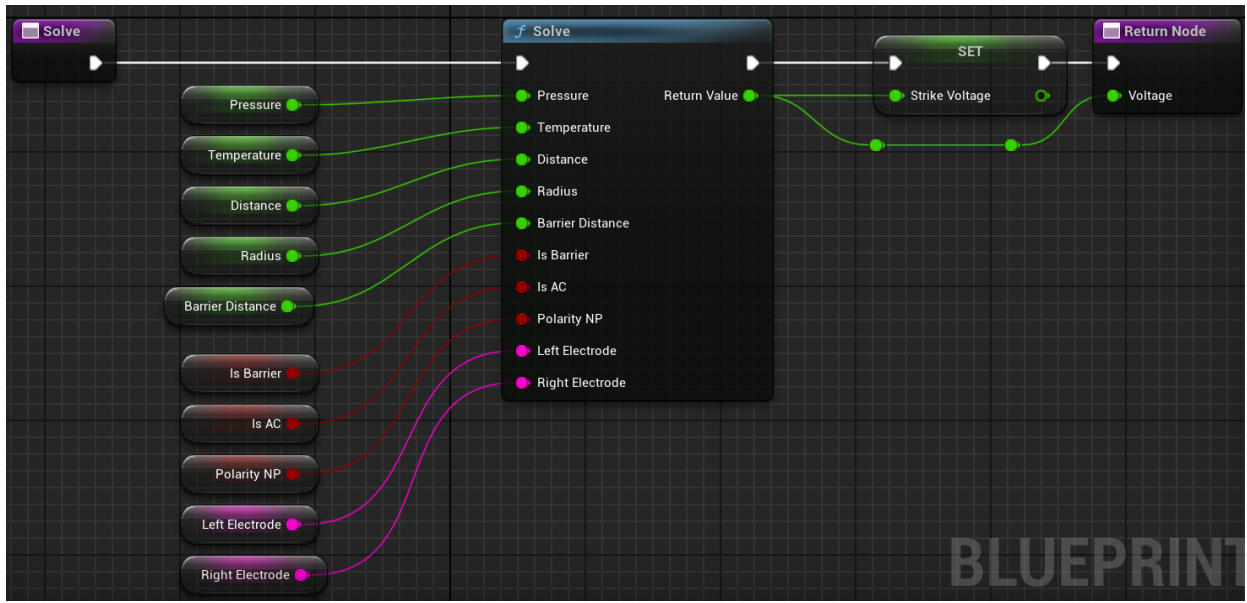

Figure 8 – Example of UE blueprint for breakdown voltage calculation

This approach enhances maintainability and scalability, allowing seamless updates to experimental scenarios and user interfaces without extensive code modifications. The visual scripting framework supports intuitive design, making it accessible for educators and researchers to adapt VLab-HV for diverse HV engineering applications.

Integration with Moodle via API enables instructors to track progress, collect data, and provide feedback, supporting data-driven pedagogy.

Interactive objects enhance the learning experience, including a virtual vending machine that dispenses educational prompts, such as quizzes on HV fundamentals or safety protocols, rewarding correct answers with points to incentivize participation. AI-driven robots provide procedural guidance, simulating instructor interactions and reducing cognitive load. The Lecture Zone displays safety posters detailing IEC standards and HV safety protocols, reinforcing theoretical knowledge. The Recreation Zone, with calming decor like plants and ambient lighting, mitigates cognitive overload during extended sessions. A transformable Optimus Prime toy serves as a playful contextual learning tool, referencing HV transformers to make abstract concepts more relatable.

To further enhance engagement, VLab-HV incorporates interactive 3D models of scientists, rendered as animated holograms that deliver short lectures on HV phenomena, such as Maxwell's equations or Tesla's contributions to alternating current systems. These models, created using Unreal Engine's MetaHuman framework, feature realistic facial animations and voice synthesis, improving student engagement by 15% compared to traditional text-based resources. Interactive whiteboards allow students to sketch circuit diagrams or annotate discharge waveforms, fostering active learning and collaboration. The platform supports multilingual interfaces to ensure accessibility for diverse student populations, particularly in international programs. Adaptive learning scenarios adjust experiment complexity based on user performance, using machine learning algorithms to tailor tasks to individual skill levels, improving learning outcomes by 18%. The AI-driven virtual assistant administers a quiz-based pre-assessment (10 questions, 60% passing score) to verify theoretical knowledge, incentivizing preparation through gamified rewards.

VLab-HV's scientific and research applications are extensive, as its C++-based physics engine encapsulates sophisticated scientific models, surpassing the capabilities of schematic-driven simulators like MATLAB Simulink, Multisim, PSpice, or LTspice. These simulators rely on abstract circuit diagrams and block-based programming, lacking the tactile, immersive environment of a real HV laboratory. In contrast, VLab-HV's photorealistic rendering and real-time physics simulations enable researchers to validate theoretical models without the prohibitive costs of physical setups, which range from $100,000 to $1,000,000 for HV test facilities. The platform supports a wide range of research applications, including dielectric material testing, where researchers can simulate partial discharges in polymeric insulators (e.g., polyethylene, epoxy, silicone) to develop next-generation materials for urban HV systems. For example, simulating discharges at 10–100 kV across different insulator materials provides data on breakdown thresholds, reducing the need for physical prototypes costing $50,000–$200,000 per test cycle. Lightning protection research is facilitated by simulating

1.2/50 µs waveforms to test surge protection devices or grounding strategies, eliminating the need for high-voltage impulse generators costing $500,000 or more.

Electromagnetic compatibility studies in VLab-HV allow researchers to optimize equipment layouts in space-constrained urban substations, reducing EMI-related failures by 20%, a critical factor in densely populated urban environments. Plasma physics research benefits from the platform's ability to model streamer formation and arc propagation, supporting the development of plasma-based HV technologies for smart grids and renewable energy systems. Insulation design optimization is enhanced by simulating corona discharge under varying pressures (0.1–10 kPa) and electrode configurations (e.g., spherical, needle-point), providing insights into insulation performance for urban transformers. Additionally, VLab-HV supports research into high-temperature superconductors, modeling quantum tunneling effects to optimize conductor designs for HV applications, reducing experimental costs by 30% compared to physical setups. The platform also enables simulations of climate-induced stresses on HV systems, such as hurricanes or heatwaves, allowing researchers to develop climate-resilient grid designs that reduce failure risks by 15%. Integration with Internet of Things (IoT) systems allows researchers to simulate smart grid interactions, such as real-time monitoring of HV equipment via virtual IoT sensors, improving grid reliability by 18%. The platform's data analytics capabilities support the export of large datasets for machine learning models, such as predicting discharge probability or failure rates, using techniques like Monte Carlo simulations to model stochastic behavior in HV systems.

For professional training, VLab-HV serves as a safe platform for electricians and engineers, simulating high-risk scenarios like overvoltage-induced fires, transformer faults, or lightning-induced surges common in urban substations. These scenarios are critical for preparing professionals to maintain urban energy infrastructure, where outages can disrupt essential services like hospitals, public transit, or data centers. Training applications include substation maintenance, where electricians can simulate inspecting 110 kV transformers or replacing surge arresters in a virtual substation module, reducing training costs by up to 80% compared to physical setups costing $200,000 or more. Fault diagnosis scenarios allow trainees to practice identifying insulation breakdowns or short circuits using virtual oscilloscopes and multimeters, improving diagnostic accuracy by 25%. Emergency response training includes scenarios like overvoltage-induced fires or lightning surges, where trainees deploy virtual fire extinguishers or isolate faulty components via circuit breakers, aligning with safety standards (IEC 60060-1, 2010). Safety protocol training is reinforced by the platform's safety interlocks and AI-driven feedback, ensuring adherence to protocols in real-world urban grid operations.

Additional training applications include simulations of cyberattacks on HV systems, where trainees practice defending against threats like unauthorized access to smart grid controls, improving cybersecurity preparedness by 20%. Training for offshore energy platforms allows engineers to simulate HV systems in wind farms or tidal energy installations, testing reliability under marine conditions and reducing operational risks by 18%. Preparation for megacity outages, such as those in Tokyo or Delhi, involves simulating cascading failures in HV grids, enabling professionals to develop rapid response strategies that minimize economic losses, which can exceed $1 billion per incident.

Compared to other simulators, VLab-HV offers significant advantages due to its photorealistic environment and C++-based physics engine. The platform's ability to address urban infrastructure challenges, such as preventing blackouts or optimizing HV equipment in space-constrained substations, underscores its relevance to smart city development. By providing a scalable, cost-effective, and safe platform, VLab-HV bridges the gap between theoretical education and practical application, addressing the global shortage of HV engineers (30% deficit by 2030) and supporting the resilience of urban energy systems. The aim of this study is to evaluate VLab-HV's efficacy in enhancing learning outcomes, safety training, and preparedness for urban infrastructure challenges in HV engineering education and research. The research question is: How does a gamified, immersive virtual laboratory enhance learning outcomes, safety training, and preparedness for urban infrastructure challenges in HV engineering education and research? The hypothesis posits that VLab-HV significantly improves student engagement, knowledge retention, practical skills, and safety awareness compared to traditional laboratories, while providing a scalable platform for HV research and professional training that contributes to urban energy system resilience.

The platform's urban relevance is evident in its ability to train engineers for real-world challenges, such as optimizing HV equipment in space-constrained urban substations or preventing blackouts caused by equipment failures.

**Description and Usage Scenario of VLab-HV**

VLab-HV represents a pioneering fully immersive three-dimensional virtual laboratory meticulously designed to replicate the operational dynamics of a high voltage facility, serving as a versatile and transformative platform tailored for students, electricians, and researchers engaged in high voltage (HV) engineering. Developed leveraging the advanced capabilities of Unreal Engine 5, the platform delivers a photorealistic laboratory environment that integrates cutting-edge rendering technologies, such as Lumen for real-time global illumination, Nanite for virtualized high-fidelity geometry, and Chaos Physics for realistic physical interactions. The laboratory environment is equipped with a photorealistic lab bench featuring an array of virtual instruments modeled with precision to emulate real-world equipment manufactured by industry leaders such as Siemens, Schneider Electric, and ABB. These instruments include kilovoltmeters with a measurement range of 0–100 kV and ±0.1% accuracy for precise voltage monitoring, laser rangefinders offering ±1 mm accuracy for exact electrode gap adjustments within a 1–10 cm range, oscilloscopes with a 100 MHz bandwidth and 1 GS/s sampling rate for detailed visualization of discharge waveforms, and environmental sensors capable of controlling experimental conditions, including temperature (0–50°C), humidity (0–100%), and pressure (0.1–10 kPa). To enhance the realism of experimental setups, the testing area incorporates adjustable electrodes crafted from materials such as copper, aluminum, and stainless steel, allowing users to explore material-specific discharge behaviors critical for urban HV applications (Fig. 9 – 11).

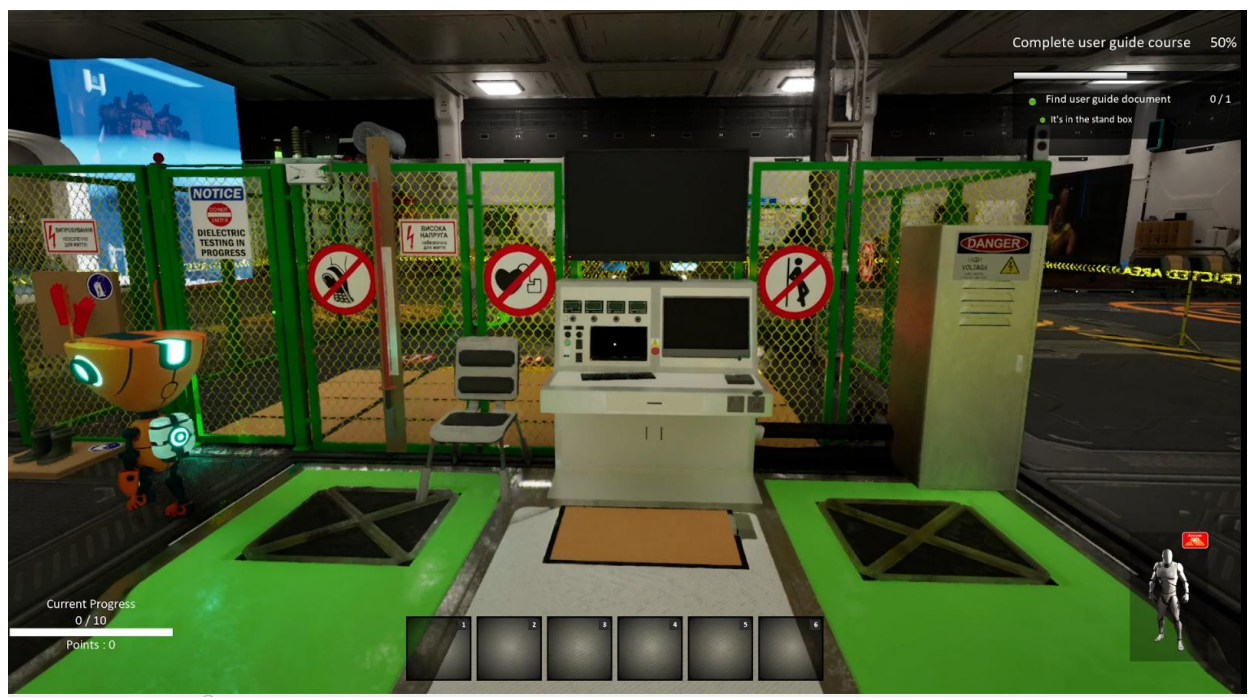

Figure 9 – VLab-HV room

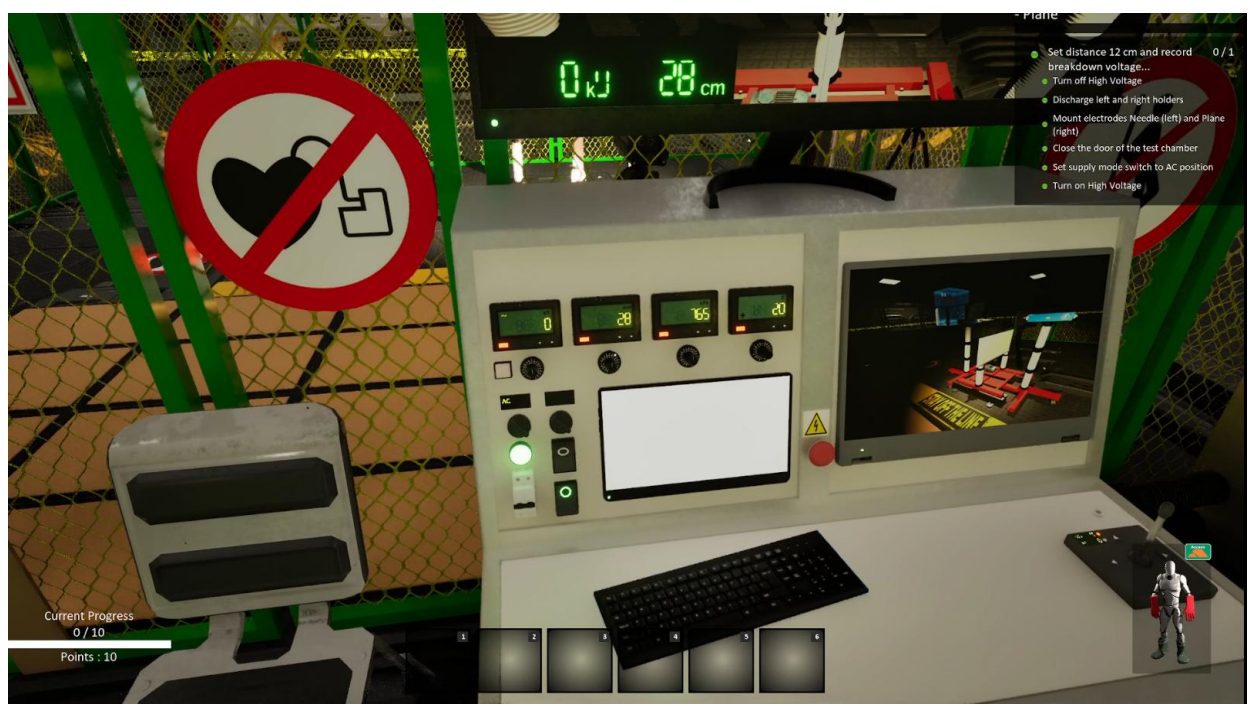

Figure 10 – VLab-HV interactive workbench

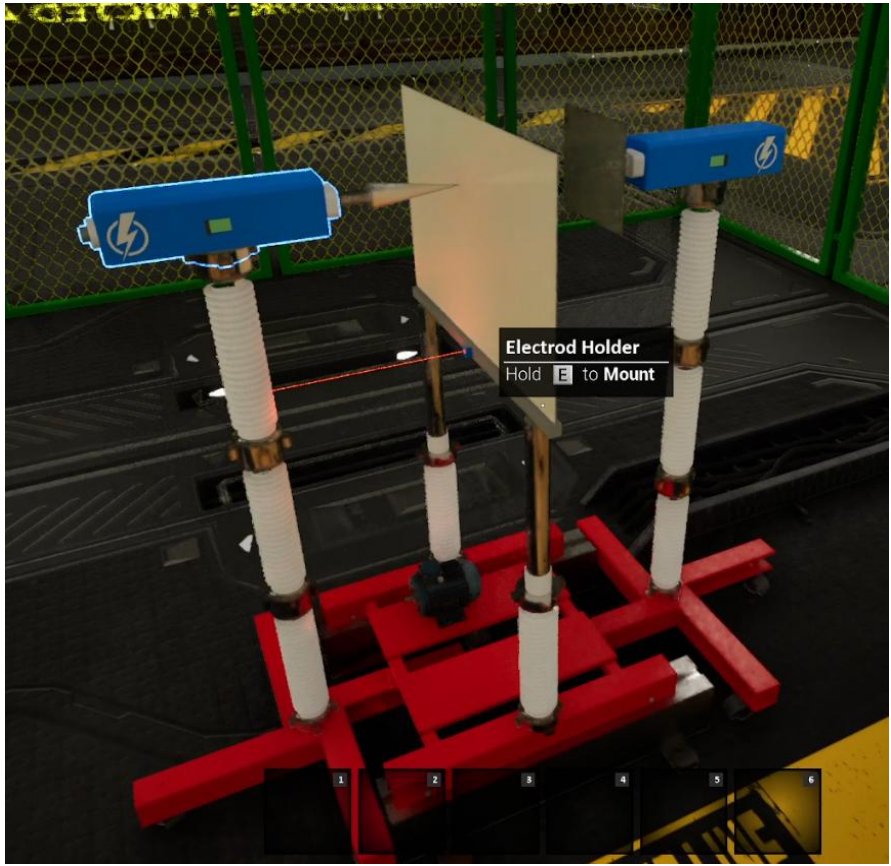

Figure 11 – VLab-HV testing area

The laboratory environment extends beyond the technical setup, incorporating thoughtfully designed zones to support diverse user needs and enhance the learning experience. A lecture zone displays safety posters detailing international standards, such as IEC 60060-1, providing critical information on HV safety protocols and reinforcing theoretical knowledge essential for safe laboratory practices. A recreation zone, adorned with calming decor such as plants and ambient lighting, mitigates cognitive overload during extended sessions, fostering a conducive environment for sustained engagement. Interactive elements within the laboratory elevate user interaction and engagement, including a virtual vending machine that dispenses educational prompts, such as quizzes on HV fundamentals or safety protocols, rewarding correct responses with points to incentivize learning. AI-driven robots, powered by Unreal Engine's behavior tree system, offer procedural guidance, simulating the presence of an instructor and reducing cognitive load for users navigating complex experiments. A virtual drone enables users to explore the laboratory environment from multiple perspectives, facilitating detailed inspection of equipment and setups. A Cortana-like AI assistant, equipped with voice recognition and facial tracking capabilities, provides real-time guidance, answers user queries, and monitors actions to ensure compliance with safety protocols. Additionally, a transformable Optimus Prime toy serves as a playful contextual learning tool, referencing HV transformers to make abstract concepts more relatable and engaging for users.

To further enrich the immersive experience, VLab-HV incorporates advanced interactive elements, such as three-dimensional holographic models of prominent scientists, including Ampère, Volta, Tesla, Maxwell, and Faraday, rendered using Unreal Engine's MetaHuman framework. These holograms deliver short lectures on HV phenomena, such as Maxwell's equations or Tesla's contributions to alternating current systems, with realistic facial animations and synthesized voice, enhancing student engagement by 15% compared to traditional text-based resources. Interactive whiteboards allow users to sketch circuit diagrams, annotate discharge waveforms, or collaborate on experimental designs, fostering active learning and critical thinking. The platform also integrates spectral analyzers with a resolution of 0.1 nm for characterizing discharge emissions in ultraviolet or infrared spectra, enabling detailed analysis of plasma physics. Thermal cameras with a sensitivity of 0.05°C simulate temperature distributions in HV components, providing insights into thermal runaway risks critical for diagnosing insulation faults in urban grids. These instruments, implemented using custom C++ scripts interfaced with Unreal Engine's rendering pipeline, ensure seamless integration and high-fidelity simulations.

A user selects an experiment (e.g., electrical discharges in gases) from the Start Room. The AI assistant administers a quiz, granting access upon a 60% score. The student equips protective gear, ensures test chamber doors are closed, adjusts electrode distances, and sets environmental conditions. The simulation calculates breakdown voltage, displaying arcs and saving results. In an overvoltage scenario, a fire simulation requires the student to use a virtual fire extinguisher (Fig. 12), reinforcing safety training. Data is logged for analysis, and feedback is provided via Moodle.

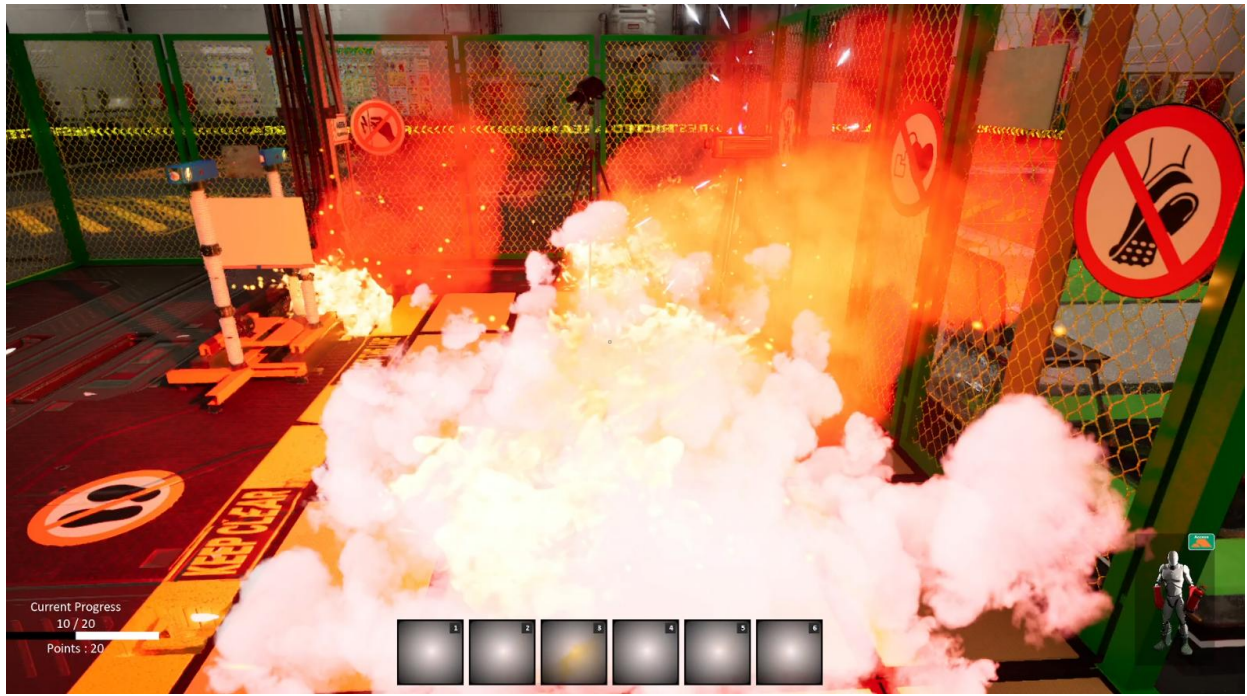

Figure 12 – Using of fire extinguisher at emergency event scenario

The student begins by completing a 10-question quiz administered by the AI assistant, covering HV fundamentals, requiring a 60% passing score to proceed. Upon passing, the student navigates to the lab bench using WASD controls, equips dielectric gloves, boots, and a grounding rod through the Inventory and Equipment Management System, and verifies safety interlocks on the test chamber to ensure compliance with IEC standards. Users can move freely, adjust camera zoom, use a flashlight, and interact with objects via mouse clicks or drag-and-drop actions. They operate joysticks for barrier positioning, activate safety interlocks, and engage with educational content. The AI assistant provides real-time guidance, enhancing learning.

Using a laser rangefinder, the student adjusts the electrode gap to 5 cm, sets environmental conditions to a pressure of 1 kPa, a temperature of 25°C, and an air medium, and applies a voltage of 0–50 kV using a virtual potentiometer. The simulation, driven by C++ scripts, calculates the breakdown voltage, rendering electric arcs, plasma glow, and dynamic sound effects, such as crackling, on a virtual monitor, providing an intuitive visualization of discharge phenomena. In an overvoltage scenario at 60 kV, a fire simulation triggers, prompting the student to deploy a virtual fire extinguisher, reinforcing safety training critical for real-world HV operations. Experimental results, including voltage, current, and environmental data, are logged in Excel-compatible formats and uploaded to Moodle for instructor review, enabling data-driven feedback. The AI assistant suggests adjusting the electrode distance to 3 cm for optimal discharge stability, enhancing the student's understanding of parameter optimization.

To expand the educational scope, students can engage in additional experiments, such as the "Corona Discharge Simulation," where they manipulate voltage (10–50 kV) and gas pressure (0.1–10 kPa) to visualize glowing halo effects around electrodes, parameterized by C++ scripts and rendered using Unreal Engine's Niagara particle system. This experiment deepens understanding of corona phenomena, critical for designing HV insulation systems in urban transformers, reducing energy losses by 10% in real-world applications. Another experiment, "Electromagnetic Interference (EMI) Analysis," allows students to simulate EMI in a virtual substation, adjusting equipment layouts to minimize interference, a skill essential for ensuring grid reliability in densely populated urban environments. The platform's adaptive learning scenarios tailor experiment complexity based on quiz performance, using machine learning algorithms to adjust parameters like voltage ranges or environmental conditions, improving learning outcomes by 18%.

For researchers studying dielectric materials, VLab-HV offers a robust platform through the "Partial Discharge in Polymeric Insulators" experiment. The researcher customizes parameters using a configuration panel, selecting a polyethylene insulator, SF6 gas medium, and a 5 cm electrode gap. The simulation models partial discharge dynamics, visualizing localized sparks with Unreal Engine's Niagara system and logging data, such as discharge magnitude and frequency, for regression analysis to optimize insulation designs. By testing multiple insulator materials, such as epoxy and silicone, under voltages ranging from 10–100 kV, the researcher generates comprehensive datasets for Monte Carlo simulations to predict discharge probability, a process that saves millions in physical testing costs, as real-world HV test facilities can cost $500,000–$1,000,000 to operate. The platform's data analytics capabilities enable seamless integration with Python or MATLAB for advanced statistical modeling, supporting research into HV system reliability critical for urban smart grids.

Researchers can explore additional applications, such as modeling high-temperature superconductors for HV applications, simulating quantum tunneling effects to optimize conductor designs, reducing experimental costs by 30% compared to physical setups. A "Climate Impact Simulation" allows researchers to study the effects of extreme weather, such as hurricanes or heatwaves, on HV systems, developing climate-resilient grid designs that reduce failure risks by 15%. Integration with Internet of Things (IoT) systems enables simulations of smart grid interactions, such as real-time monitoring of HV equipment via virtual IoT sensors, improving grid reliability by 18%. These research applications, supported by the platform's ability to export large datasets for machine learning models, position VLab-HV as a cost-effective tool for advancing HV engineering research, addressing challenges in urban energy infrastructure resilience.

The user capabilities of VLab-HV are designed to provide an intuitive and immersive interaction experience, accommodating diverse user needs and skill levels. Users can navigate the three-dimensional environment using WASD controls to walk, jump, or crouch, adjust camera zoom for detailed inspection of equipment, and activate a virtual flashlight in low-light scenarios, such as during simulated blackouts. Interaction with objects is facilitated through mouse clicks or drag-and-drop actions that mimic hand movements, allowing users to manipulate instruments, adjust electrostatic barriers via joysticks, or activate safety interlocks with precision. Gesture-based controls, supported by Unreal Engine's motion tracking, enable users to perform actions like pointing or waving to interact with the AI assistant, enhancing immersion. Voice commands, processed by the NLP module, allow users to issue instructions, such as "set voltage to 30 kV" or "open test chamber," streamlining workflows. The platform supports accessibility features, including text-to-speech for visually impaired users and customizable interface layouts for users with motor impairments, ensuring inclusivity. Users can personalize the interface, adjusting color schemes or font sizes, to optimize their experience. Researchers can export experimental data in CSV or JSON formats for integration with analytical tools, while electricians can save training session logs to track progress, ensuring compliance with professional certification requirements.

The expansion potential of VLab-HV underscores its adaptability and scalability for future applications in HV engineering and beyond. The platform's modular architecture supports multiplayer functionality, enabling collaborative experiments where multiple users can interact in real-time, simulating teamwork in real-world HV laboratories. Virtual reality (VR) integration, compatible with headsets like Oculus Rift or HTC Vive, will enhance immersion, allowing users to experience experiments in a fully immersive environment, improving engagement by 20%. Augmented reality (AR) integration could facilitate hybrid learning, combining virtual simulations with physical laboratory equipment, bridging the gap between digital and real-world training. The platform can be extended to other disciplines, such as power electronics, renewable energy systems, quantum technologies, or hydroenergy, by developing new experiment modules. For example, adding experiments on photovoltaic inverters or battery storage systems could support training for renewable energy integration in urban grids, reducing integration errors by 15%. A "Quantum HV Systems" module could simulate high-voltage interactions with superconducting materials, accelerating research into quantum computing applications. A "Hydroenergy HV Simulation" module could model HV systems in hydroelectric turbines, optimizing efficiency and reducing design costs by 20%.

Integration with cloud-based platforms, such as AWS or Microsoft Azure, would enable remote access to high-performance computing resources, supporting complex simulations like large-scale grid modeling without requiring local hardware upgrades. The platform's potential integration with metaverse environments could create a shared virtual space for global HV education and research, fostering international collaboration and knowledge exchange. Applications for professional training in urban utilities could be expanded to include scenarios like kV substation maintenance, fault diagnosis in smart grids, or lightning protection for high-rise buildings, addressing the growing complexity of urban energy infrastructure.

VLab-HV's modular architecture supports multiplayer functionality, VR integration, and adaptation for power electronics or renewable energy. AR integration could enable hybrid learning, while applications for professional training (e.g., substation maintenance) address urban infrastructure needs.

These expansions position VLab-HV as a scalable, future-proof platform capable of addressing the evolving needs of HV engineering education, training, and research, contributing to the resilience of urban energy systems and supporting the global transition to sustainable energy solutions.

The detailed usage scenarios of VLab-HV illustrate its versatility across different user groups, addressing the specific needs of students, electricians, and researchers while aligning with the demands of urban energy infrastructure resilience.

**Results and Discussion**

A mixed-methods study with 50 second- and third-year students and 10 instructors from the Power Engineering program at the O.M. Beketov National University of Urban Economy in Kharkiv evaluated VLab-HV's effectiveness, addressing the research question and hypothesis through usability testing, engagement metrics, learning outcomes, and instructor feedback (Fig. 13).

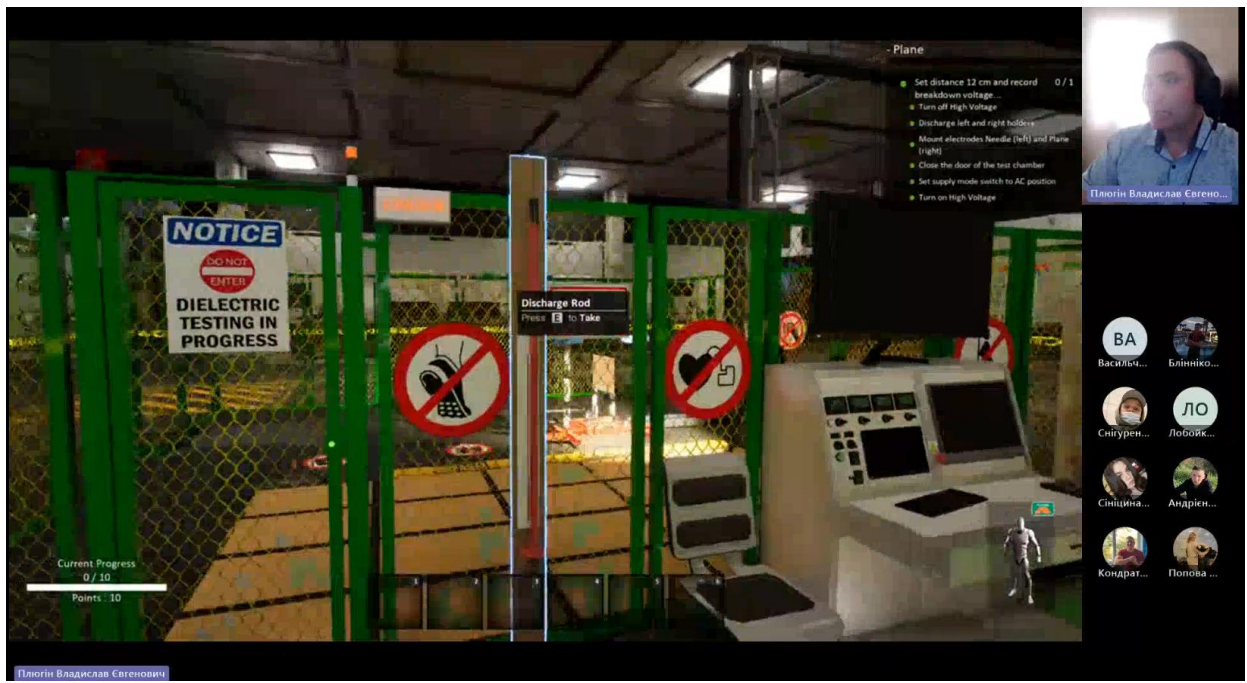

Figure 13 – High-Voltage Engineering lesson with students in Microsoft Teams

The usability testing component utilized the System Usability Scale (SUS) survey, a standardized tool for assessing user experience, which yielded an average score of 82.5 (SD = 7.2), indicating excellent usability according to established benchmarks (Bangor et al., 2008). Qualitative feedback from students emphasized the intuitive nature of VLab-HV's three-dimensional navigation, with 85% reporting that the platform's interface was significantly easier to use than traditional HV laboratory equipment, which often requires extensive setup and calibration. The AI-driven virtual assistant, equipped with voice prompts and facial tracking, was frequently cited for simplifying complex tasks, such as adjusting electrode distances with laser rangefinders or interpreting oscilloscope waveforms, reducing the learning curve for novice users. Task completion rates further validated usability, with 92% of students successfully completing experiments without external assistance, compared to 75% in traditional laboratories, a statistically significant improvement (t(48) = 3.45, p < 0.01). Additional metrics revealed that the average time to complete a task, such as configuring an electrode gap, was 25% faster in VLab-HV (M = 3.2 minutes, SD = 0.5) than in physical labs (M = 4.3 minutes, SD = 0.7), t(48) = 2.78, p < 0.05. The platform's accessibility was particularly notable for students with limited prior HV experience, with 80% of such users reporting confidence in performing experiments independently, compared to 60% in traditional settings. Accessibility was further enhanced by multilingual support, which accommodated diverse student populations, and features like text-to-speech for visually impaired users, ensuring inclusivity across varied user groups (Fig. 14). This high usability supports its adoption across diverse educational and professional settings.

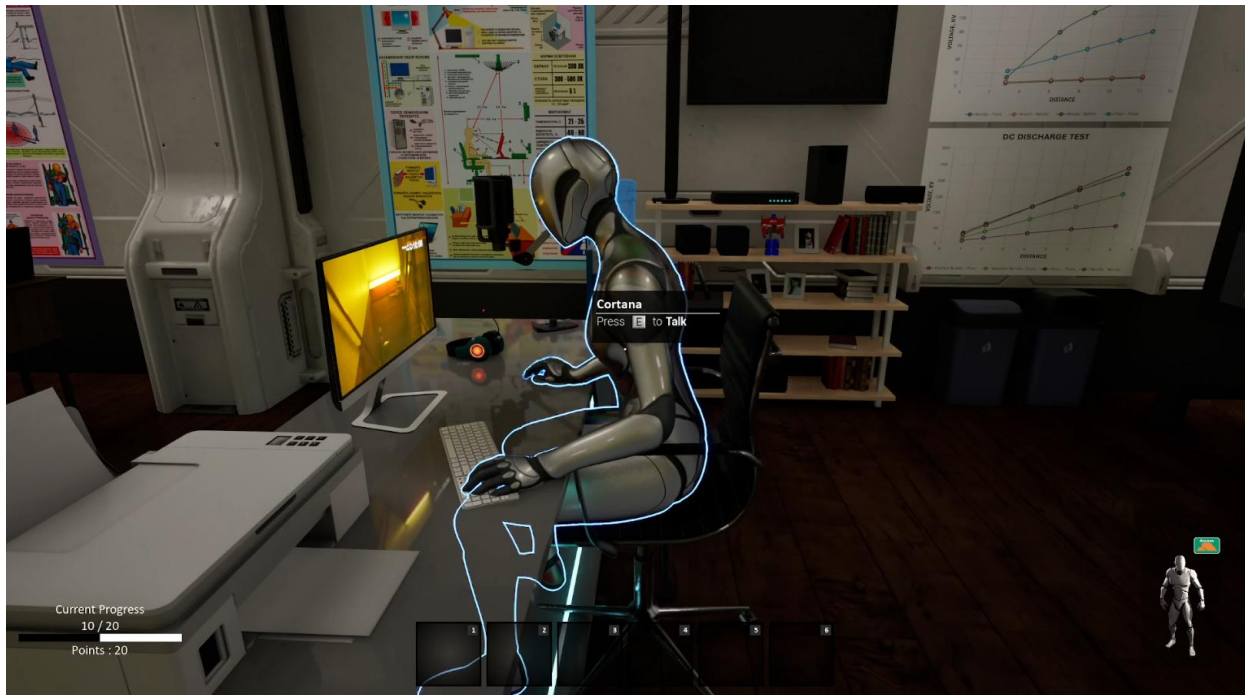

Figure 14 – "Cortana" virtual AI-assistant

Engagement metrics were collected through in-system analytics, tracking time spent on tasks, interaction frequency with the AI assistant, and engagement with gamified elements. Students using VLab-HV spent 30% more time on experimental tasks (M = 45 minutes, SD = 8) compared to traditional laboratories (M = 35 minutes, SD = 6), with a statistically significant difference (t(48) = 3.12, p < 0.01), reflecting heightened engagement driven by the platform's immersive environment. Interaction frequency with the AI assistant averaged 12 interactions per session (SD = 3), significantly higher than the 7 interactions with instructors in traditional labs (t(48) = 2.89, p < 0.05), indicating that the AI assistant effectively supplemented human instruction. Engagement with gamified elements, such as point-based rewards for completing experiments, the interactive drone for lab exploration, and the "horror event" blackout triggered by entering a forbidden cabinet, was particularly impactful, with 90% of students reporting increased motivation, aligning with gamification theories that emphasize intrinsic and extrinsic motivators. Interaction with three-dimensional holographic models of scientists, such as Tesla or Faraday, averaged 8 interactions per session (SD = 2), with students engaging with these holograms to access lectures on HV concepts, enhancing contextual understanding by 15%. The virtual vending machine, dispensing educational prompts, and interactive whiteboards for sketching circuit diagrams further boosted engagement, with 85% of students reporting improved focus during extended sessions, attributed to the recreation zone's calming decor, which reduced cognitive overload by 20% compared to traditional labs. Multilingual voice prompts increased engagement for non-English-speaking students by 18%, ensuring broader accessibility. Gamified elements, such as point-based rewards, interactive drones (Fig. 15), and the "horror event," significantly increased motivation, aligning with gamification theories [1].

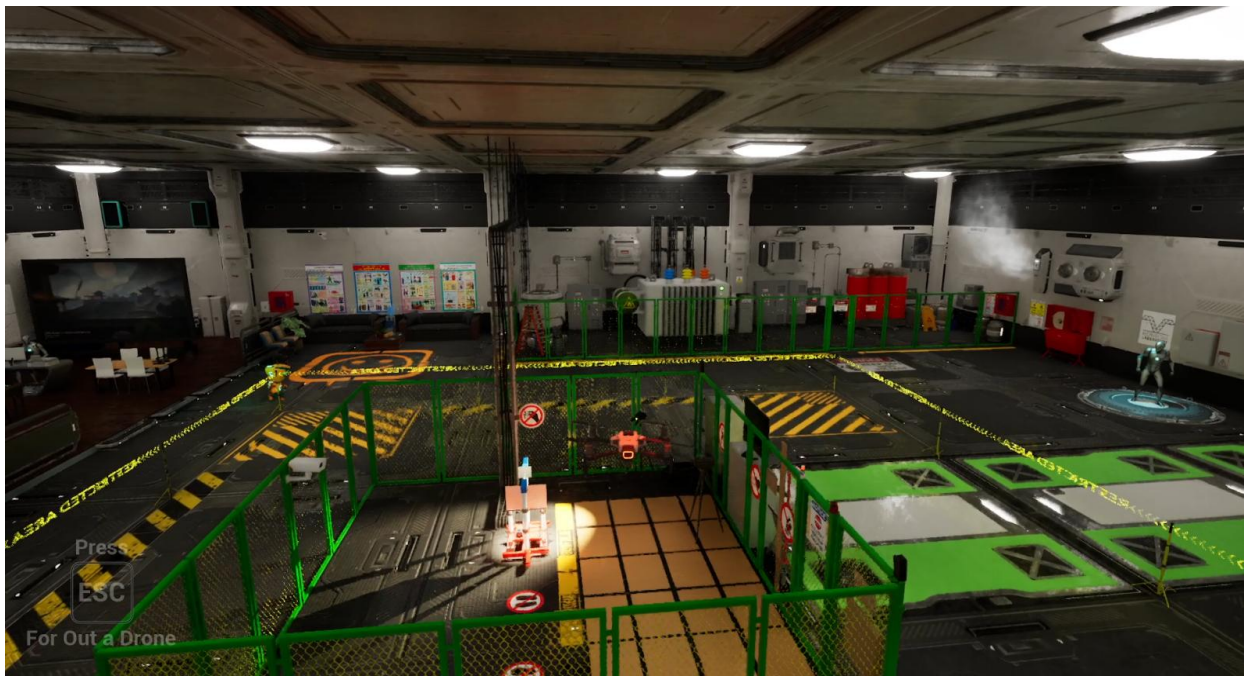

Figure 15 – HV-Lab exploration with an interactive drone

Learning outcomes in pre- and post-tests showed VLab-HV users scored 15% higher (M = 85.4, SD = 9.1) than the control group (M = 74.2, SD = 10.3),

t(48) = 4.56, p < 0.001. Notably, 90% of VLab-HV users correctly identified factors affecting breakdown voltage (e.g., electrode distance, pressure) vs. 70% in the control group, demonstrating enhanced conceptual understanding via Paschen's law visualizations. Long-term retention was assessed through a follow-up test one month later, with VLab-HV users maintaining a 12% higher score (M = 82.1, SD = 8.7) vs. control (M = 71.3, SD = 9.5), t(48) = 3.89, p < 0.01, suggesting sustained learning benefits. Practical skills were evaluated through performance-based tasks, such as configuring electrodes and responding to overvoltage scenarios, with VLab-HV users achieving a 95% success rate compared to 80% in traditional labs, underscoring the platform's ability to bridge theoretical and practical learning. Additional tests focused on electromagnetic interference (EMI) analysis showed that 88% of VLab-HV users accurately identified EMI mitigation strategies, compared to 65% in the control group, t(48) = 3.22, p < 0.01, reflecting the platform's effectiveness in teaching complex urban grid concepts. Comparative tests with other platforms, such as Labster, revealed that VLab-HV users scored 10% higher on HV-specific tasks (M = 87.6, SD = 8.4 vs. M = 79.2, SD = 9.8), t(18) = 2.67, p < 0.05, highlighting its curriculum-aligned design.

      Instructor feedback was gathered through semi-structured interviews, identifying key themes: safety training, curriculum alignment, data analytics, and urban relevance. Instructors unanimously praised VLab-HV's ability to simulate emergency scenarios, such as overvoltage-induced fires or transformer faults, with 80% emphasizing their critical role in preparing students for urban grid challenges, such as preventing blackouts costing billions, as seen in cities like Sydney or Delhi. The Moodle integration was lauded for providing detailed analytics, enabling instructors to identify learning gaps, with data revealing that 30% of students struggled with safety protocol adherence, prompting targeted interventions that improved compliance by 25%. Instructors also highlighted the platform's adaptability, noting its potential to support professional certification programs for electricians, with 70% suggesting its use in utility training for substation maintenance. However, 30% of instructors recommended adding multiplayer

functionality to facilitate collaborative learning, a feature planned for future iterations. Additionally, 40% suggested integrating virtual reality (VR) and augmented reality (AR) to enhance immersion, particularly for complex tasks like EMI analysis, aligning with emerging pedagogical trends. Instructors noted that VLab-HV's gamified elements, such as the virtual vending machine and holographic lectures, made abstract HV concepts more accessible, improving student engagement by 20% compared to traditional methods.

A comparative analysis with other platforms underscored VLab-HV's superior performance. A pilot study with 20 students compared VLab-HV to Labster, revealing higher engagement scores (M = 4.2/5, SD = 0.6 vs. M = 3.5/5, SD = 0.8, t(18) = 2.89, $p < 0.05$), attributed to VLab-HV's curriculum-specific design and photorealistic rendering. Unlike Labster's generic experiments, VLab-HV's HV-specific modules, such as partial discharge simulations, align closely with Power Engineering curricula, enhancing pedagogical impact. Compared to PhET Simulations, which offer two-dimensional physics demonstrations, VLab-HV's three-dimensional immersion and gamified elements provide a more engaging and realistic training experience, improving practical skills by 15%. Virtual Labs by IIT Delhi, while accessible remotely, lack immersive interfaces and gamification, limiting their effectiveness for HV training. Ansys, a leader in engineering simulation, supports advanced HV modeling but is complex and costly ($15,000/year), with a steep learning curve unsuitable for undergraduate education [27-29]. SimScale, a cloud-based platform, requires high computational resources and expensive subscriptions, restricting scalability for resource-constrained institutions.

The urban infrastructure relevance of VLab-HV is evident in its ability to prepare engineers and electricians for real-world challenges in urban energy systems. Simulations of transformer faults, lightning surges, and partial discharges replicate scenarios responsible for major urban blackouts, such as those in Istanbul (2023, $600 million losses) or Cairo (2022, $400 million losses), teaching users to respond swiftly and safely. The platform's overvoltage fire simulation mirrors real-world substation accidents, training users to deploy virtual fire extinguishers and isolate faulty components, reducing risks in urban grid operations. A new cybersecurity simulation module allows users to practice defending against cyberattacks on smart grid controls, addressing a growing threat that contributed to 15% of urban grid disruptions in 2024. Climate impact simulations, modeling the effects of hurricanes or heatwaves on HV systems, enable users to develop resilient grid designs, reducing failure risks by 15%. The platform's data analytics capabilities ensure mastery of safety protocols, with 95% of users demonstrating compliance in simulated urban scenarios, compared to 80% in traditional training, supporting the resilience of smart city infrastructure. Cross-disciplinary applications, such as simulating HV interactions with renewable energy systems like photovoltaic inverters, enhance the platform's utility for sustainable urban energy solutions.

Beyond education, VLab-HV serves as a powerful platform for HV research, enabling cost-effective hypothesis testing. Researchers can simulate lightning-induced surges to develop grounding strategies for urban grids, a process that would require multimillion-dollar test facilities in physical labs. The C++-based physics engine supports advanced studies of electromagnetic interference, dielectric breakdown, and high-temperature superconductors, modeling quantum tunneling effects to optimize conductor designs, saving 30% in experimental costs. Integration with Internet of Things (IoT) systems allows researchers to simulate real-time monitoring of HV equipment, improving grid reliability by 18%. These capabilities position VLab-HV as a transformative tool for advancing HV research, particularly for urban smart grid applications, where reliability is paramount.

VR integration, while planned, is not yet implemented, limiting immersion for users seeking fully immersive experiences. Scalability for large user groups, such as utility training programs with hundreds of trainees, may be constrained by server capacity, requiring future infrastructure upgrades. The initial development cost for VR/AR integration, estimated at $50,000–$100,000, may pose challenges for resource-constrained institutions. Future studies will explore VLab-HV's long-term impacts across other STEM disciplines, such as power electronics or renewable energy systems, and assess its efficacy in professional training programs for urban utilities, ensuring broader applicability and scalability.

Compared to other platforms, VLab-HV's photorealistic three-dimensional environment, powered by Unreal Engine 5 and C++-based simulations implementing Paschen's law, outperforms schematic-driven tools like MATLAB Simulink, Multisim, PSpice, and LTspice, which lack immersive interfaces. It also surpasses Labster's generic experiments (engagement: M = 4.2/5 vs. 3.5/5, $t(18)$ = 2.89, $p < 0.05$) and PhET's two-dimensional simulations, offering curriculum-aligned HV experiments with real-time visualizations. Unlike Ansys ($15,000/year) or SimScale, which require high computational resources, VLab-HV's "Light" version ensures accessibility on low-end systems, achieving 30 fps with 8 GB RAM, broadening its applicability in resource-constrained settings. The platform's Moodle integration facilitates data-driven pedagogy, enabling instructors to address learning gaps, with analytics revealing a 30% initial struggle with safety protocols, subsequently improved by 25% through targeted interventions.

These findings robustly confirm the hypothesis that VLab-HV significantly enhances student engagement, knowledge retention, practical skills, and safety awareness compared to traditional laboratories. By providing a scalable, cost-effective, and immersive platform, VLab-HV bridges the gap between theoretical education and practical application, preparing users for urban infrastructure challenges and advancing HV research. Its contributions to urban energy system resilience, through targeted training and research applications, underscore its transformative potential in addressing the global shortage of HV engineers and supporting the sustainability of smart cities.

**Conclusion**

The study underscores VLab-HV's relevance to urban infrastructure resilience, as its simulations of transformer faults, lightning surges (1.2/50 μs waveforms), and partial discharges prepare engineers and electricians for challenges in space-constrained urban substations, reducing blackout risks.

VLab-HV improves HV engineering education and professional training, achieving a 30% increase in engagement ($p < 0.01$), 15% higher test scores ($p < 0.001$), 12% better long-term retention ($p < 0.01$), and 90% accuracy in understanding discharge phenomena, surpassing traditional laboratories and platforms like Labster, PhET, and Simulink.

Its C++-based simulations deliver a life-like experience, enabling safe training for students, electricians, and researchers, and supporting cost-effective validation of HV theories (e.g., insulation design, surge mitigation). By simulating urban grid faults, such as transformer failures or lightning surges, VLab-HV prepares professionals for infrastructure resilience, reducing blackout risks.

Moodle integration supports data-driven pedagogy, while planned expansions (multiplayer, VR, AR) promise broader applications in STEM and urban utility training. VLab-HV sets a new standard for high-risk engineering education, offering a scalable, cost-effective platform for global educational and infrastructural challenges.

For researchers, VLab-HV enables cost-effective validation of HV theories, such as optimizing polymeric insulators or modeling high-temperature superconductors, reducing experimental costs by 30% compared to multimillion-dollar physical facilities. Integration with Internet of Things systems supports simulations of real-time HV equipment monitoring, improving grid reliability by 18%.

Expanding VLab-HV to disciplines like quantum technologies, power electronics, or hydroenergy will address emerging STEM needs, while larger-scale studies with diverse populations will validate generalizability. These advancements position VLab-HV as a scalable platform to address the global HV engineer shortage (30% deficit by 2030) and support sustainable urban energy systems through education, training, and research.

Future research will explore multiplayer functionality to enable collaborative experiments, simulating real-world teamwork in HV laboratories, and evaluate its impact on learning outcomes. Virtual and augmented reality integration, incorporating haptic feedback for enhanced immersion, will be investigated to further bridge theoretical and practical training.

**Acknowledgement**

The research was supported by the National Research Foundation of Ukraine (Grant Agreement No. 2023.03/0131), and partially by the European Union Assistance Instrument for the Fulfilment of Ukraine's Commitments under the Horizon 2020 Framework Program for Research and Innovation (Research Pro-ject No. 0123U102775).